\documentclass[11pt,letterpaper]{article}
\usepackage[normalem]{ulem}
\usepackage{color}
\usepackage{graphicx}
\usepackage{subfig}
\usepackage{bbm,amsmath,amssymb,mathrsfs,verbatim,epsfig,wasysym,slashed}
\usepackage{jheppub}

\newcommand{\Ref}[1]{Ref.~\cite{#1}}
\newcommand{\Refs}[1]{Refs.~\cite{#1}}
\newcommand{\Sec}[1]{Sec.~\ref{#1}}
\newcommand{\App}[1]{App.~\ref{#1}}
\newcommand{\Fig}[1]{Fig.~\ref{#1}}
\newcommand{\Figs}[2]{Figs.~\ref{#1} and \ref{#2}}
\newcommand{\Tab}[1]{Table~\ref{#1}}
\newcommand{\Eq}[1]{Eq.~(\ref{#1})}
\newcommand{\HC}{\text{h.c.}}
\newcommand{\vis}{\text{vis}}
\newcommand{\hid}{\text{hid}}
\newcommand{\tot}{\text{tot}}

\newcommand{\bs}{\boldsymbol}

\renewcommand{\Re}{\text{Re}}
\renewcommand{\Im}{\text{Im}}

\newcommand{\MPl}{M_\text{Pl}}

\def\lsim{\mathrel{\rlap{\lower4pt\hbox{\hskip1pt$\sim$}}
    \raise1pt\hbox{$<$}}}
\def\gsim{\mathrel{\rlap{\lower4pt\hbox{\hskip1pt$\sim$}}
    \raise1pt\hbox{$>$}}} 
\newcommand{\vev}[1]{ \left\langle {#1} \right\rangle }

\newcommand{\kev}{{\rm keV}}
\newcommand{\mev}{{\rm MeV}}
\newcommand{\gev}{{\rm GeV}}
\newcommand{\tev}{{\rm TeV}}

\newcommand{\hc}[1]{#1^{\dagger}}

\newcommand{\be}{\begin{eqnarray}}
\newcommand{\ee}{\end{eqnarray}}

\newcommand{\parfrac}[2]{\left(\frac{#1}{#2}\right)}

\newcommand{\order}{\mathcal{O}}

\newcommand{\ssoftmass}[1]{\widetilde{m}^2_{#1}}

\newcommand{\gravitino}{\widetilde{G}}

\newcommand{ \chiEaten}{\chi_{\rm eaten}}

\newcommand{ \chiUneaten}{\chi_{\rm uneaten}}

\newcommand{ \psGld}{\zeta}
\newcommand{ \chiVis}{\chi_{\rm vis}}
\newcommand{ \chiHid}{\chi_{\rm hid}}

\newcommand{\sPot}{\mathcal{W}}
\newcommand{\mgrav}{m_{3/2}}
\newcommand{\kahler}{\mathcal{K}}

\title{Visible Supersymmetry Breaking and an Invisible Higgs}

\author{Daniele Bertolini,}
\author{Keith Rehermann,}
\author{and Jesse Thaler}
\affiliation{Center for Theoretical Physics,\\
 Massachusetts Institute of Technology, Cambridge, MA 02139, U.S.A.}
\emailAdd{danbert@mit.edu}
\emailAdd{krmann@mit.edu}
\emailAdd{jthaler@mit.edu}
\abstract{If there are multiple hidden sectors which independently break supersymmetry, then the spectrum will contain multiple goldstini.  In this paper, we explore the possibility that the visible sector might also break supersymmetry, giving rise to an additional pseudo-goldstino.  By the standard lore, visible sector supersymmetry breaking is phenomenologically excluded by the supertrace sum rule, but this sum rule is relaxed with multiple supersymmetry breaking.  However, we find that visible sector supersymmetry breaking is still phenomenologically disfavored, not because of a sum rule, but because the visible sector pseudo-goldstino is generically overproduced in the early universe.  A way to avoid this cosmological bound is to ensure that an $R$ symmetry is preserved in the visible sector up to supergravity effects.   A key expectation of this $R$-symmetric case is that the Higgs boson will dominantly decay invisibly at the LHC.}


\begin{document}

\hfill MIT-CTP 4320

\maketitle

\section{Introduction}
\label{sec:Intro}

Spontaneously broken supersymmetry (SUSY) is an appealing solution to the gauge hierarchy problem.  A crucial question for SUSY phenomenology is how SUSY breaking is communicated to the Supersymmetric Standard Model (SSM).  The well-known supertrace sum rule prohibits SUSY breaking from occurring directly in the SSM through renormalizable tree-level interactions \cite{Haber:1993wf,Martin:1997ns,Luty:2005sn}.  This observation has led to the standard two-sector paradigm, where a hidden sector is responsible for SUSY breaking, and the visible sector (i.e.\ the SSM) feels SUSY breaking indirectly via messenger fields.

Recently, it has been argued that the standard two-sector paradigm may be too restrictive, as there could exist multiple hidden sectors which independently break SUSY \cite{Cheung:2010mc}.   A striking signature of this proposal is that if SUSY is broken by $N$ independent sectors, then there is a corresponding multiplicity of ``goldstini''.   One linear combination is eaten to form the longitudinal component of the gravitino, while the remaining $N-1$ modes remain in the spectrum as uneaten goldstini.\footnote{The phenomenological implications of goldstini have been studied in detail in \Refs{Cheung:2010mc,Cheung:2010qf,Craig:2010yf,McCullough:2010wf,Cheng:2010mw,Izawa:2011hi,Argurio:2011hs,Thaler:2011me,Cheung:2011jq}.  The idea of pseudo-goldstinos first appeared in the context of brane-worlds in \Ref{Benakli:2007zza}.}

\begin{figure}[t]
\begin{center}
\subfloat[]{\includegraphics[width=0.45\textwidth]{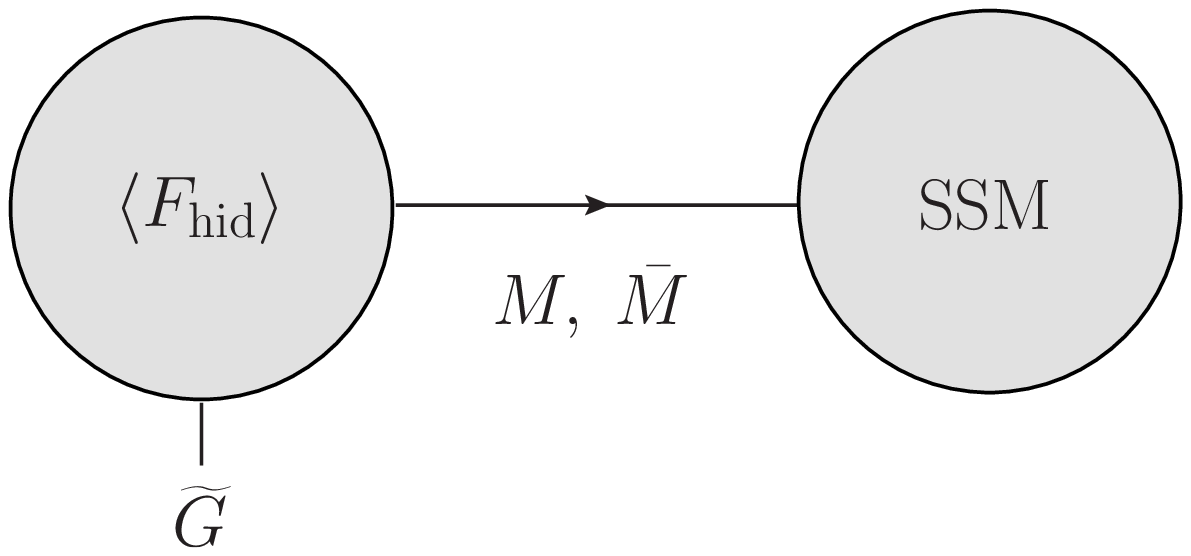}} $\qquad$
\subfloat[]{\includegraphics[width=0.45\textwidth]{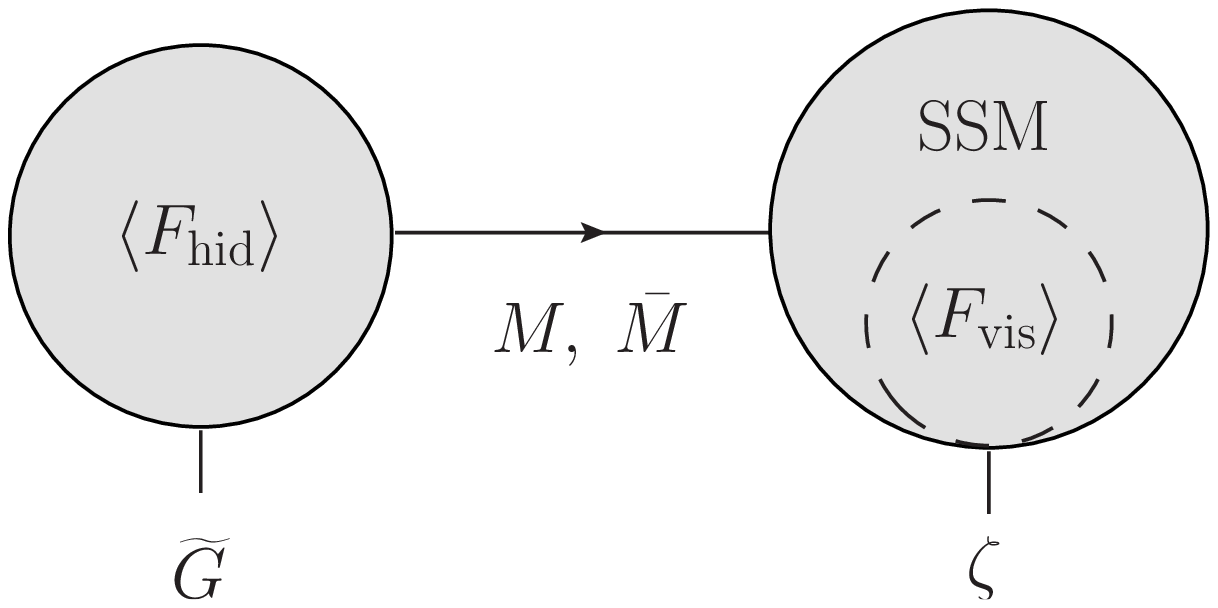}}
\caption{Left:  In the standard paradigm, SUSY is broken in the hidden sector and communicated to the visible sector via messenger fields.  The hidden sector goldstino is eaten by gravitino $\gravitino$.  Right:  SUSY can also be broken in the visible sector, giving rise to a visible pseudo-goldstino $\psGld$.  To evade the supertrace sum rule, there must be additional SSM soft masses mediated from the hidden sector.   While the standard assumption is that this mediation is $R$-violating, we will also consider $R$-symmetric mediation.}
\label{fig:genericSetup}
\end{center}
\end{figure}

Motivated by the possibility of multiple SUSY breaking, in this paper we reexamine the usual assumption that SUSY cannot be broken in the visible sector.  As long as there are one or more hidden sectors contributing to SSM soft masses, then the supertrace sum rule constraint does not apply, and SUSY can indeed be broken in the SSM at tree-level.  Analogous to \Ref{Cheung:2010mc} and previously discussed in \Ref{Izawa:2011hi}, this leads to an uneaten goldstino in the visible sector.  Unlike \Ref{Cheung:2010mc}, the uneaten goldstino mixes with SSM fields, but despite this mixing, there is still a light mass eigenstate which we refer to as a pseudo-goldstino.  For concreteness, we study the simplest example of visible sector SUSY breaking from \Ref{Izawa:2011hi}, where the minimal $R$-symmetric SSM \cite{Kribs:2007ac} is extended to allow for $F$-term breaking.  The generic setup we envision is shown in \Fig{fig:genericSetup}. 

The phenomenology of the pseudo-goldstino depends sensitively on its mass, which in turn depends on how hidden sector SUSY breaking is mediated to the SSM.  In the usual case with $R$-violating SSM soft parameters, the pseudo-goldstino has a mass of $\order(10~\mev-1~\gev)$, which implies significant cosmological constraints.   Thus, the standard lore that SUSY cannot be broken in the ($R$-violating) SSM is essentially correct, albeit not because of the supertrace sum rule but because of pseudo-goldstino overproduction in the early universe.  That said, there are small corners of parameter space with healthy pseudo-goldstino cosmology.

On the other hand, if the mediation mechanism preserves an $R$-symmetry, then the pseudo-goldstino will only get a mass from ($R$-violating) supergravity (SUGRA) effects proportional to $\mgrav$.   Thus, if the gravitino $\gravitino$ is sufficiently light (as expected to avoid the cosmological gravitino problem \cite{Pagels:1981ke}), then the pseudo-goldstino is also cosmologically safe.  There are interesting collider implications for the $R$-symmetric limit, since the light pseudo-goldstino $\psGld$ is typically accompanied by a light pseudo-sgoldstino $\phi$.  Intriguingly, we will find that in much of parameter space, the physical Higgs boson $h^0$ dominantly decays invisibly as $h^0 \rightarrow \phi \phi \rightarrow \psGld \gravitino \psGld \gravitino$, affecting Higgs discovery prospects at the LHC.  

The remainder of this paper is organized as follows.   In \Sec{sec:model}, we describe the simplest model of visible sector SUSY breaking, and discuss $R$-violating and $R$-symmetric mediation.  In \Sec{sec:MassWidth}, we calculate the mass and lifetime of the pseudo-goldstino for both types of mediation.  We discuss cosmological constraints in \Sec{sec:Cosmo} and LHC signatures in \Sec{sec:LHC}.  We conclude in \Sec{sec:conclude}, leaving calculational details to the appendices.

\section{Breaking Supersymmetry in the Visible Sector}
\label{sec:model}

There are a variety of models which break SUSY at tree-level, generalizing the familiar O'Raifeartaigh model.  To truly have SUSY breaking in the visible sector, SUSY breaking must involve SSM multiplets in some way.   Because gauge quantum numbers restrict the types of interactions possible, it is most natural for SUSY breaking to involve just the Higgs multiplets of the SSM. 

In this section, we review the minimal model of visible sector SUSY breaking previously studied in \Ref{Izawa:2011hi}, and identify the pseudo-goldstino mode.  We then introduce the effects of the hidden sector, and explain why the pseudo-goldstino remains light even in the presence of SSM soft masses.  Though we will confine our discussion to the minimal model, more general SUSY breaking scenarios are likely to share much of the same phenomenology, since our analysis is largely based on the symmetries of the low energy theory.  The key ingredient is a pseudo-goldstino of $R$-charge 1 that can mix with higgsino and gaugino modes after electroweak symmetry breaking.

\subsection{Visible Sector SUSY Breaking}
\label{subsec:VisSusy}

The minimal model of visible sector SUSY breaking is \cite{Izawa:2011hi}
\begin{equation}
\label{eq:W}   
\mathcal{W}=\mathcal{W}_{\rm Yukawa}+X(\lambda H_u H_d-\kappa)+\mu_u H_u R_u+\mu_d H_d R_d,
\end{equation}
where the standard Yukawa interactions are
\begin{equation}\label{Wyuk}   
\mathcal{W}_{\rm Yukawa}= y_u QH_u U^c + y_d Q H_d D^c + y_e L H_d E^c.
\end{equation}
Like the minimal $R$-symmetric SSM, there are two sets of Higgs doublets $H_{u,d}$ and $R_{u,d}$ with vector-like mass terms.  Like the next-to-minimal SSM, there is a gauge singlet field $X$.  This superpotential respects a $U(1)_R$ symmetry with the charge assignments in \Tab{tab1}.  We will not dwell on the ultraviolet (UV) origin of the mass parameters in \Eq{eq:W}, though such mass terms are often dynamically generated in composite Higgs theories \cite{Harnik:2003rs,Chang:2004db,Delgado:2005fq,Craig:2011ev,Csaki:2011xn}.\footnote{We note that the $\mu$-terms in \Eq{eq:W} are consistent with being generated by the Giudice-Masiero mechanism \cite{1988PhLB..206..480G}, however $\kappa$ is not.} 

\begin{table}[t]
\begin{center}
\begin{tabular}{ | c | c | }
\hline
\rule[-2mm]{0mm}{.6cm}
Superfield & $U(1)_R$\\ \hline \hline
\rule[-2mm]{0mm}{.6cm}
$H_u$, $H_d$ & 0\\ \hline
\rule[-2mm]{0mm}{.6cm}
$Q$, $U^c$, $D^c$, $L$, $E^c$& 1\\ \hline
\rule[-2mm]{0mm}{.6cm}
$R_u$, $R_d$, $X$ & 2\\ \hline
\end{tabular}
\end{center} 
\caption{The $R$-charge assignments for the minimal model of visible sector SUSY breaking.}
\label{tab1}
\end{table}

In the absence of SSM soft masses, \Eq{eq:W} spontaneously breaks SUSY.  The electromagnetically neutral part of the tree-level scalar potential is:
\begin{equation}
V_{\vis}=V_F+V_D,
\end{equation}
where
\begin{equation}
V_F=\left|\lambda h_u^0h_d^0-\kappa\right|^2+\left|\lambda xh_d^0+\mu_u r_u^0\right|^2+\left|\lambda xh_u^0+\mu_d r_d^0\right|^2+\mu_u^2\left|h_u^0\right|^2+\mu_d^2\left|h_d^0\right|^2,
\end{equation}
\begin{equation}
V_D=\frac{1}{8}(g^2+g'^2)\left(\left|h_u^0\right|^2-\left|h_d^0\right|^2+\left|r_d^0\right|^2-\left|r_u^0\right|^2\right)^2,
\end{equation}
and we use a notation where lower-case characters stand for the scalar components of the corresponding superfield.

Since there is no way to simultaneously satisfy all of the $F$-term equations of motion, SUSY is spontaneously broken.  At tree-level, there are two types of minima in terms of $(x,h_u^0,h_d^0,r_u^0,r_d^0)$: \begin{itemize}
\item SUSY breaking but gauge-preserving minima: $\mbox{Min}_1=(\langle x\rangle,0,0,0,0)$;
\item SUSY breaking and gauge-breaking minima: $\mbox{Min}_2=(\langle x\rangle, \langle h_u^0\rangle,\langle h_d^0\rangle,\langle r_u^0\rangle,\langle r_d^0\rangle)$.
\end{itemize}
Formulas for the gauge-breaking minima appear in \Ref{Izawa:2011hi}.  In both cases, the $x$ flat direction is lifted by quantum corrections and the vacuum expectation value (vev) $\langle x\rangle$ is stabilized at zero. Since $\langle r_u^0\rangle$ and $\langle r_d^0\rangle$ are proportional to $\langle x\rangle$, both kinds of minima preserve the $R$-symmetry of \Tab{tab1}. 

Notice that the $R$-symmetry predicts three massless neutral fermions at tree level.  This is because only two linear combinations of the the $R$-charge $+1$ fermions ($\widetilde{x}$, $\widetilde{r}_u^0$, $\widetilde{r}_d^0$, $\widetilde{B}$, $\widetilde{W}_3$) can marry the two $R$-charge $-1$ fermions ($\widetilde{h}_u^0$, $\widetilde{h}_d^0$) to make $R$-invariant Dirac masses.  Therefore, three linear combinations of the $R$-charge +1 fermions must be massless.  Spontaneous SUSY breaking ensures that one of the three massless states is the visible sector goldstino:
\begin{equation}\label{chivis}
\chiVis \simeq \langle F_X\rangle\widetilde{x}+\langle F_{R_u}\rangle\widetilde{r}_u^0+\langle F_{R_d}\rangle\widetilde{r}_d^0+\langle D_Y \rangle\widetilde{B} +\langle D_3\rangle\widetilde{W_3},
\end{equation}  
where $\langle F_{R_u}\rangle$, $\langle F_{R_d}\rangle$, $\langle D_Y\rangle$, and $\langle D_3\rangle$ are only non-vanishing  for the gauge-breaking minima.  Note that because of the preserved $R$ symmetry, the $H_{u,d}$ multiplets do not have $F$-components in the vacuum.  The other two massless fermions correspond roughly to the bino and wino of the SSM.

\subsection{Hidden Sector SUSY Breaking}
\label{subsec:HidSusy}

In order to evade the supertrace sum rule, \Eq{eq:W} must be augmented by hidden sector SUSY breaking.  Regardless of the details of the hidden sector dynamics, this implies a hidden sector goldstino $\chiHid$ in addition to the visible sector goldstino $\chiVis$.  One linear combination is eaten via the super-Higgs mechanism to form the longitudinal component of the gravitino
\begin{equation}\label{chihid}
\chiEaten = \frac{\langle F_{\vis}\rangle \chiVis + \langle F_{\hid}\rangle\chiHid}{F},
\end{equation}
where $F_\vis \equiv \sqrt{V_\vis}$ and $F_\hid \equiv \sqrt{V_\hid}$ are the respective contributions to SUSY breaking from the visible and hidden sectors, and the total amount of SUSY breaking is
\be
F \equiv \sqrt{\vev{F_{\vis}}^2+\vev{F_{\hid}}^2}.
\ee
In the limit where the visible and hidden sectors are completely sequestered, the orthogonal combination of fermions
\begin{equation}
\chiUneaten = \frac{\langle F_{\hid}\rangle \chiVis - \langle F_{\vis}\rangle\chiHid}{F}
\end{equation}
is an uneaten goldstino.  After zeroing the cosmological constant, the gravitino mass is
\be
\mgrav = \frac{F}{\sqrt{3} \MPl},
\ee
and the uneaten goldstino gets a mass proportional to $\mgrav$ from SUGRA effects \cite{Cheung:2010mc,Cheung:2011jq}. 

Taking $m_{3/2}$ to be much smaller than the weak scale, the fermionic spectrum contains two light states, the gravitino and the uneaten goldstino.  For the rest of the paper, we assume $\vev{F_{\vis}} \ll \vev{F_{\hid}}$ such that $\chiEaten \simeq \chiHid$ and $\chiUneaten \simeq \chiVis$ in the sequestered limit.  

\subsection{Soft Terms}

To generate SSM soft terms, the hidden and visible sectors cannot be completely sequestered and must interact via messengers.  The leading phenomenological effect of the messenger sector can be captured by the resulting SSM soft terms.  The soft terms consistent with SSM charge assignments but not necessarily with the $R$-symmetry in \Tab{tab1} are
\begin{equation}\label{eq:soft}
\begin{split}
\mathcal{L}_{\mbox{\scriptsize{soft}}}=
&-\frac{1}{2}M_1\widetilde{B}\widetilde{B} -\frac{1}{2}M_2\widetilde{W}\widetilde{W} -\frac{1}{2}M_3\widetilde{g}\widetilde{g}+ \HC \\
&-A_h xh_uh_d-B_u h_ur_u-B_d h_dr_d-Tx + \HC \\
&-\widetilde{m}^2_{H_u}\left|h_u\right|^2- \widetilde{m}^2_{H_d}\left|h_d\right|^2-\widetilde{m}^2_{R_u}\left|r_u\right|^2-\widetilde{m}^2_{R_d}\left|r_d\right|^2-\widetilde{m}^2_{X}\left|x\right|^2\\
&+\mathcal{L}_{\mbox{\scriptsize{soft}}}^{\mbox{\scriptsize{Matter}}},\\
\end{split}
\end{equation}
where $\mathcal{L}_{\mbox{\scriptsize{soft}}}^{\mbox{\scriptsize{Matter}}}$ stands for SSM matter field soft terms. For simplicity we have elided
soft terms that do not have any counterpart in the superpotential \Eq{eq:W} and off-diagonal scalar soft masses.\footnote{Such terms do not arise if SUSY breaking is mediated to the visible sector solely through a superfield of $R$-charge 2, where the $R$-symmetry is spontaneously broken by its vev.  This is indeed the case, for instance, in gauge mediation and anomaly mediation.   More generally, although additional soft terms like $B_r r_u r_d$ or $B_h h_u h_d$ do modify the vacuum structure, the mass of the goldstino is not substantially modified, as explained by the persistent zero mode argument in \Sec{eq:whyLight}.} If the mediation respects an $R$-symmetry, then only the soft masses $\widetilde{m}^2$ are generated.\footnote{Majorana masses for the gauginos violate the $R$-symmetry, necessitating new field content to achieve Dirac gaugino  masses.  We will discuss this in more detail in \Sec{sec:RSmass}.}

In the presence of SSM soft terms, the $H_{u,d}$ multiplet can now obtain non-zero $F$-components, deforming the visible sector goldstino away from $\chiVis$:
\be
\label{eq:modChiVis}
\chiVis' \sim \chiVis + \vev{F_{H_u}} \widetilde{h}_u^0 +\vev{F_{H_d}} \widetilde{h}_d^0.
\ee
However, since the soft terms affect the vacuum structure, there is no guarantee that $\chiVis'$ will even be a mass eigenstate,\footnote{In addition, the messenger sector generically introduces new fermionic mass terms that mix the hidden sector and visible sector goldstinos.  In the $\vev{F_{\vis}} \ll \vev{F_{\hid}}$ limit, we can safely ignore such effects.} but we will see that there is still a light fermion in the spectrum.  

\subsection{A GeV-scale Pseudo-Goldstino?}
\label{eq:whyLight}

There are two facts which conspire to ensure a light fermion in the visible sector spectrum.  This state is generically different from \Eq{eq:modChiVis}, so we will refer to it as a pseudo-goldstino and denote it by $\psGld$.

\begin{itemize}

\item \textbf{Persistent Zero Mode in Wess-Zumino Models}: In the absence of gauge interactions, the visible sector superpotential in \Eq{eq:W} is an example of a (renormalizable) Wess-Zumino model.  With a minimal K\"ahler potential, the fermionic mass matrix is
\be
\mathcal{M}_{ab}(\phi)=\frac{\partial^2W}{\partial\phi_a\partial\phi_b},
\ee
and because \Eq{eq:W} spontaneously breaks SUSY, $\det \mathcal{M}_{ab}(\vev{\phi}) = 0$ in the vacuum.  Moreover, for Wess-Zumino models that spontaneously break SUSY, $\det \mathcal{M}_{ab}(\phi) = 0$ for \emph{arbitrary} scalar field configurations.\footnote{This result is reasonably well-known in the literature, though much of it unpublished.  See \Ref{Visser:1984aq} for a straightforward argument using the Witten index \cite{Witten:1982df}.}

Now consider adding SSM soft masses.  At tree-level and in the absence of gauge interactions, the only effect of adding \Eq{eq:soft} is to change the vacuum configuration of the visible sector fields.  However, since $\det \mathcal{M}_{ab}(\phi) = 0$ for all field configurations, there is guaranteed to be a massless fermion at tree-level.  Thus, the pseudo-goldstino can only get a tree-level mass through gauge interactions, namely through mixing with the gauginos.   We will see that this mixing angle is quite small, thus the leading pseudo-goldstino mass is loop suppressed.

\item \textbf{R Symmetry}:  As discussed in \Sec{subsec:VisSusy}, the visible sector $R$-symmetry implies three massless fermions.  Thus, the pseudo-goldstino mass is proportional to the degree of $R$-violation.  If the mediation preserves an $R$-symmetry, then at minimum, the pseudo-goldstino will get a mass from SUGRA effects proportional to $m_{3/2}$.
In the usual case that $R$-symmetry is broken by SSM soft masses, the pseudo-goldstino mass will depend on the $R$-violating gaugino masses, $A$-terms, $B$-terms, and $x$ tadpole.  As already mentioned, the tree-level effect is small because it is proportional to the small goldstino/gaugino mixing angle.  The $R$-violating scalar soft terms contribute to the pseudo-goldstino mass only at loop level.\footnote{In addition, the $R$-violating scalar soft terms themselves are often suppressed (notably in gauge mediation), leading to an additional suppression of the pseudo-goldstino mass.}

\end{itemize}

To illustrate these points, consider a hidden sector field $S$ with $R$-charge 2 and the visible sector field $X$ also of $R$-charge 2.  In the $\vev{F_{\vis}} \ll \vev{F_{\hid}}$ limit, we can apply the arguments above to understand the mass of the visible sector fermion in $X$.  Integrating out the messenger sector at loop level leads to non-minimal K\"ahler couplings between the hidden and visible sectors.  The K\"ahler operator
\be
\label{eq:RviolOper}
\frac{c_1}{\Lambda} (S + S^\dagger) (X^\dagger X)
\ee
is an example of an $R$-violating operator which contributes to SSM soft terms.  However, this term does not contain a fermion mass for $X$ so it does not evade the first point.\footnote{This operator appears to induce a Dirac mass between the fermion in $S$ and the fermion in $X$, but this mass must vanish in the vacuum to have a massless true goldstino.}  The $R$-symmetric K\"ahler operator
\be
\label{eq:FermContainOper}
\frac{c_2}{\Lambda^2} (X^\dagger X)^2
\ee
does contain a fermion mass for $X$ proportional to $\vev{x}$, but it cannot induce a mass unless the $R$-symmetry is broken by another operator to give a non-zero value of $\vev{x}$.  Therefore, only when both types of operators are present can a pseudo-goldstino mass be generated.

To summarize, even after coupling the visible sector to a hidden source of SUSY breaking, a light pseudo-goldstino persists as a remnant of the original visible SUSY breaking dynamics.   Its tree-level mass is suppressed because it is only induced by small mixings with the gauginos.  At one loop, its mass is protected by the $R$-symmetry.  These two effects imply that the pseudo-goldstino mass is typically a loop factor below the scale of $R$-violation in the SSM soft parameters, putting it in the (cosmologically dangerous) mass range $\order(10~\mev-1~\gev)$.  For $R$-symmetric mediation, the mass is suppressed and proportional to $m_{3/2}$ (and cosmologically safe for $\mgrav \ll 1~\kev$).  Since the above arguments are based mainly on $R$-symmetry and SUSY, one expects them to hold  on quite general grounds independent of the details of the visible SUSY breaking dynamics.

The cosmological bounds in \Sec{sec:Cosmo} on the $R$-violating scenario would be weakened if the pseudo-goldstino could be made heavier than a few GeV.  In principle, and at the price of tuning electroweak symmetry breaking, the loop-induced mass could be raised above naive estimates by increasing the size of the $R$-violating soft parameters, though arbitrarily large soft terms will spoil electroweak symmetry breaking.  We could try to increase the size of $R$ violation by considering visible sector SUSY breaking which spontaneously breaks $R$ \cite{Komargodski:2009jf}, but by the Wess-Zumino zero mode argument, this $R$ violation would feed into the pseudo-goldstino mass only at loop level.  Finally, we note that the mass of the light fermion can be raised with an operator $\sPot \supset m X^2$.  Of course, with such an operator, SUSY is no longer broken in the visible sector, and there is no sense in which the light fermionic state can be referred to as a pseudo-goldstino.

\section{Properties of the Pseudo-Goldstino}
\label{sec:MassWidth}

As discussed in the previous section, the properties of the pseudo-goldstino are strongly influenced by the SUSY breaking mediation mechanism.  In the case of $R$-violating mediation, there are significant one-loop corrections to pseudo-goldstino mass.  Conversely, if the mediation is $R$-symmetric, the mass of the pseudo-goldstino is proportional to $m_{3/2}$ but typically lighter than the gravitino.  We begin by calculating the mass and width of the pseudo-goldstino in the presence of $R$-violation, and then study the $R$-symmetric case.

\subsection{Mass with R Violation}
\label{sec:RVmass}

For arbitrary vevs of the neutral scalars, the tree-level neutralino mass matrix in the basis
\begin{equation}
 {\bs \psi} =\left(\widetilde{x},\widetilde{h}_u^0,\widetilde{h}_d^0,\widetilde{r}_u^0,\widetilde{r}_d^0,\widetilde{B},\widetilde{W}_3\right)
\end{equation}
is
\begin{equation}
\mathcal{M}=\left(
\begin{array}{ccccccc}
0 & \lambda\langle h_d^0\rangle & \lambda\langle h_u^0\rangle & 0 & 0  & 0&0\\
 \lambda\langle h_d^0\rangle & 0 & \lambda\langle x\rangle & \mu_u & 0 & \frac{g'\langle h_u^0\rangle}{\sqrt{2}}& -\frac{g\langle h_u^0\rangle}{\sqrt{2}}\\
\lambda\langle h_u^0\rangle & \lambda\langle x\rangle & 0 & 0 & \mu_d &  -\frac{g'\langle h_d^0\rangle}{\sqrt{2}}& \frac{g\langle h_d^0\rangle}{\sqrt{2}}\\
0 &\mu_u & 0 & 0 & 0 &  -\frac{g'\langle r_u^0\rangle}{\sqrt{2}}& \frac{g\langle r_u^0\rangle}{\sqrt{2}}\\
0 & 0 & \mu_d & 0 & 0 &  \frac{g'\langle r_d^0\rangle}{\sqrt{2}}& -\frac{g\langle r_d^0\rangle}{\sqrt{2}}\\
0 & \frac{g'\langle h_u^0\rangle}{\sqrt{2}}&-\frac{g'\langle h_d^0\rangle}{\sqrt{2}}& -\frac{g'\langle r_u^0\rangle}{\sqrt{2}} &  \frac{g'\langle r_d^0\rangle}{\sqrt{2}} &  M_1&0\\ 
0 & -\frac{g\langle h_u^0\rangle}{\sqrt{2}}&\frac{g\langle h_d^0\rangle}{\sqrt{2}}& \frac{g\langle r_u^0\rangle}{\sqrt{2}} & -\frac{g\langle r_d^0\rangle}{\sqrt{2}} & 0 &M_2\\ 
\end{array} 
\right).
\end{equation}

As argued in \Sec{eq:whyLight}, the tree-level pseudo-goldstino mass is induced only by mixing with the gauginos. Expanding in the gauge couplings, the first-order (unnormalized) mass eigenstate is
\be
\label{eq:firstordermass}
\psGld: \quad \left(1,0,0,- \frac{\lambda \vev{h^0_d}}{\mu_u},-  \frac{ \lambda \vev{h^0_u}}{\mu_d}, - \frac{g'}{\sqrt{2}}\frac{\lambda \vev{r'}}{M_1},  \frac{g}{\sqrt{2}} \frac{\lambda \vev{r'}}{M_2}\right) + \mathcal{O}(g^2),
\ee
where we have defined the $R$-charge 2 combination
\be
\label{eq:rprimedef}
r' \equiv \frac{ h^0_u}{\mu_d} r^0_d -\frac{h^0_d}{\mu_u} r^0_u.
\ee
The tree-level mass of the pseudo-goldstino is
\be
\label{eq:mtree}
m^{\rm tree}_{\psGld} =  \frac{\lambda^2}{2} \vev{r'}^2 \left(\frac{g'^2}{M_1} + \frac{g^2}{M_2}  \right) + \mathcal{O}(g^2).
\ee
After solving for the vacuum configuration, we find that for typical weak-scale values for the soft masses and superpotential parameters $(\order(100~\gev))$, the pseudo-goldstino mass is $m^{\rm tree}_{\psGld} \simeq \order(1-10~\mev)$. In particular, as long as all of the Higgs sector soft parameters have a similar scale, then there is a cancellation in \Eq{eq:rprimedef} which yields a small value of $\vev{r'}$, and thus a small pseudo-goldstino mass.\footnote{It is possible to increase the tree-level pseudo-goldstino mass to $\order(1~\gev)$ by imposing a large up/down hierarchy on the Higgs sector soft parameters. That said, this larger mass is still constrained by the cosmological bounds in \Sec{sec:Cosmo}.}

\begin{figure}[t]
\begin{center}
\includegraphics[scale=0.7]{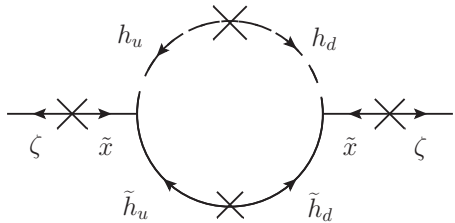}
\caption{Estimate of the loop correction to the pseudo-goldstino mass.  Fermion and scalar insertions come from the superpotential \Eq{eq:W} and from the $A_h$ term in \Eq{eq:soft}.  The fermion insertion is $\lambda\langle x\rangle$ and the scalar insertion is $2\lambda\kappa-A_h\langle x\rangle-\lambda^2\langle h_u^0\rangle\langle h_d^0\rangle$.  The full set of diagrams appear in \Fig{fig:feyn1}. }
\label{diagnaive}
\end{center}
\end{figure}

Given the small tree-level effect, we need to take into account loop corrections.  At this order, the contribution from gauginos is small, and the pseudo-goldstino can be treated as a linear combination of $\widetilde{x}$, $\widetilde{r}_u^0$, and $\widetilde{r}_d^0$.  Throughout this paper, we will use the notation
\be
\Theta_{g,m}
\ee
to denote the mixing angle between the gauge eigenstate $g$ and the mass eigenstate $m$.  The diagram shown in \Fig{diagnaive} gives a naive estimate for the one loop correction:  
\begin{equation}
\label{eq:oneLoopMassEstimate}
\delta m^{\rm loop}_\psGld \approx \frac{2\lambda^2\Theta_{\widetilde{x},\psGld}^2}{16\pi^2}\frac{\lambda\langle x\rangle\left( 2\lambda\kappa-A_h\langle x\rangle-\lambda^2\langle h_u^0\rangle\langle h_d^0\rangle\right)}{m_{\rm eff}^2}.
\end{equation}
Here, $\lambda^2/(16\pi^2)$ is a loop factor and the $2$ accounts for both neutral and charged particles in the loop.  The fermion mass insertion $\lambda\langle x\rangle$ and the scalar mass insertion $2\lambda\kappa-A_h\langle x\rangle-\lambda^2\langle h_u^0\rangle\langle h_d^0\rangle$ come from \Eq{eq:W} and \Eq{eq:soft}, and $m_{\rm eff}$ is the characteristic mass scale for the particles in the loop. 

\begin{figure}[t]
\begin{centering}
\subfloat[]{\label{fig:mass1} \includegraphics[width=0.45\textwidth]{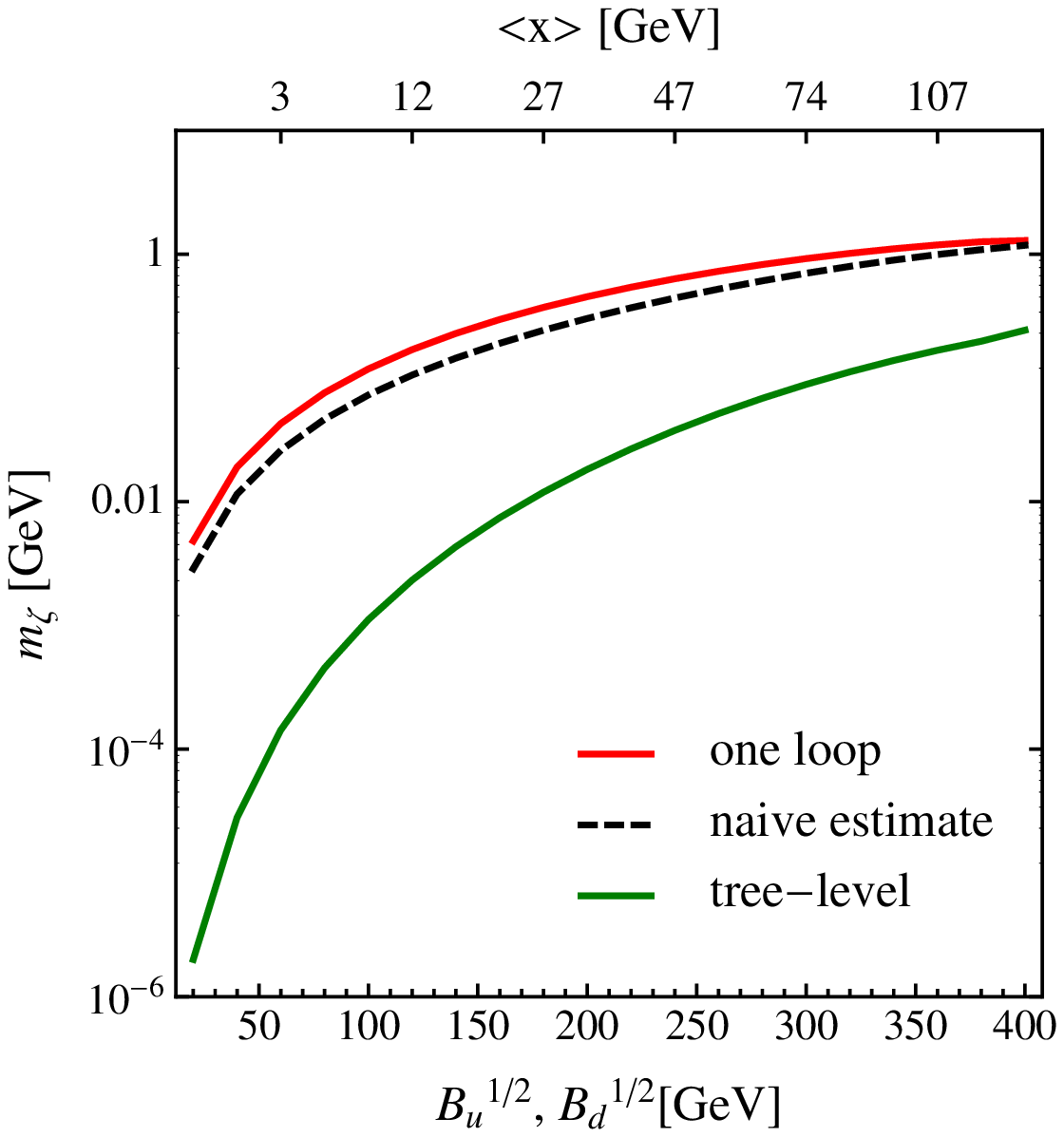}}\hfill
\subfloat[]{\label{fig:mass2} \includegraphics[width=0.45\textwidth]{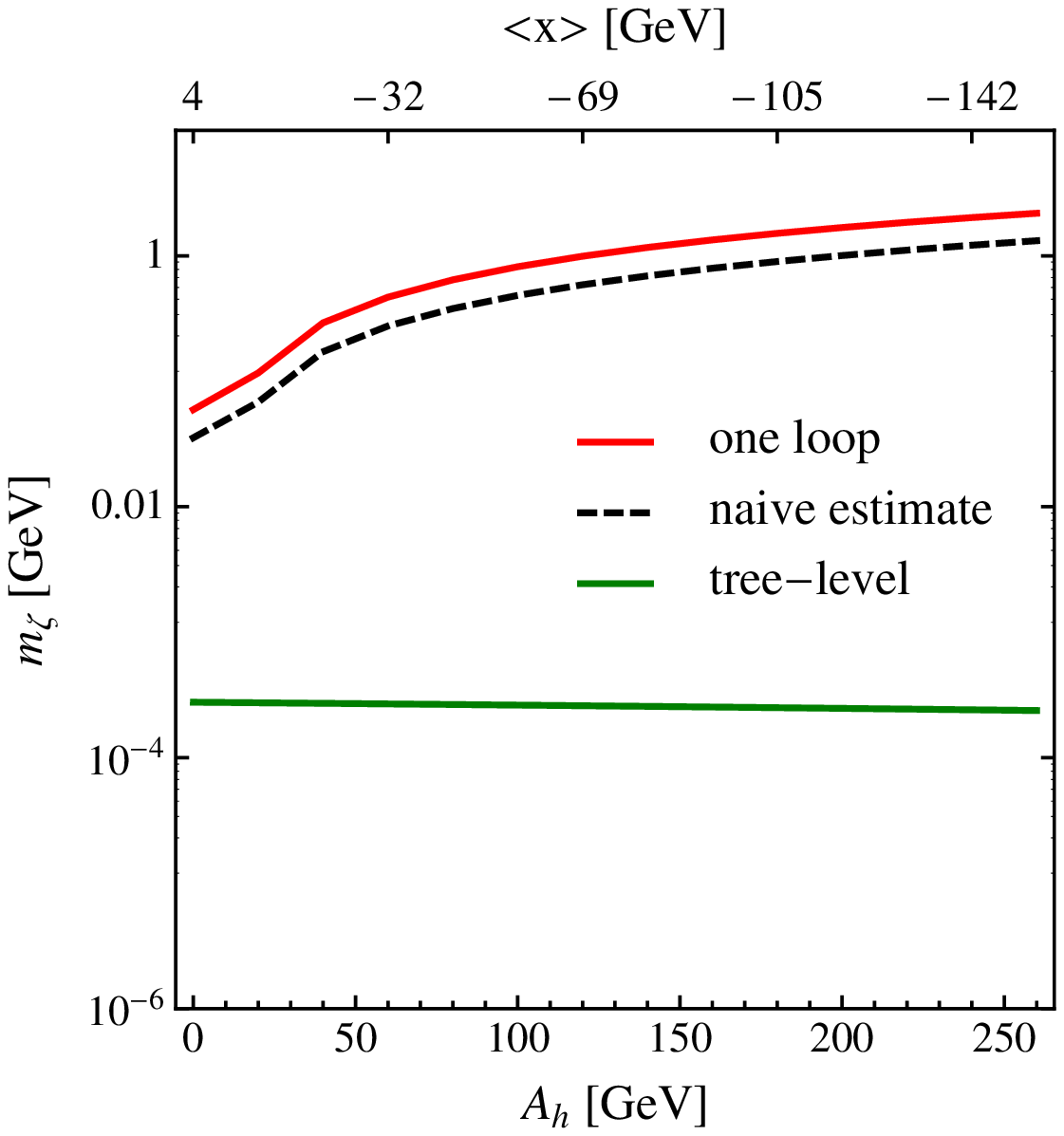}}
\caption{Left: mass of the pseudo-goldstino as a function of  $B_u^{1/2}=B_d^{1/2}$. For concreteness, we fix $\widetilde{m}_{H_u}^2=\widetilde{m}_{H_d}^2=\widetilde{m}_{R_u}^2=\widetilde{m}_{R_d}^2=(600~\gev)^2$, $\mu_u=200~\gev$, $\mu_d=300~\gev$, $M_1=100~\gev$, $M_2=150~\gev$, and $\lambda=1$.  The value of $\kappa$ is chosen to obtain the correct value of $m_Z$, and all other soft parameters are set to zero.  The red line indicates the mass including one loop corrections and the dashed line is the naive estimate according to \Eq{eq:oneLoopMassEstimate} with $m_{\rm eff}^2=\widetilde{m}_{H_u}^2+\widetilde{m}_{H_d}^2+\mu_u^2+\mu_d^2$.  For comparison, the green line shows the small tree-level contribution.  The value of $\vev{x}$ is shown as a reference, since this vev controls the mass according to \Eq{eq:oneLoopMassEstimate}.  For $R$-breaking soft terms around the weak scale, the mass falls in the range $\order(10\;\mev -1~\gev)$.  Right: mass of the pseudo-goldstino as a function of  $A_h$ with $B_u=B_d=(70~\gev)^2$ and all other soft parameters as in the left figure.} 
\label{fig:PGmass}
\end{centering}
\end{figure}

A full calculation of the one-loop pseudo-goldstino mass appears in \App{sec:oneLoop}, but we can estimate the size of the effect from \Eq{eq:oneLoopMassEstimate}.  If we take $\kappa \gg \langle h_u^0\rangle\langle h_d^0\rangle, A_h\langle x\rangle$ we find
\begin{align}
\label{approx2}
 \delta m^{\rm loop}_\psGld &\approx \frac{\lambda^4\Theta_{\widetilde{x}, \psGld}^2}{4\pi^2}\frac{\langle x\rangle\kappa}{m_{\rm eff}^2} \\ &\approx 100 \; \mev \parfrac{\lambda}{1.0}^4 \parfrac{\Theta_{\widetilde{x}, \psGld}}{0.7}^2 \parfrac{\vev{x}}{35 \; \gev} \parfrac{\kappa}{(100~\gev)^2}\parfrac{300 ~\gev}{m_{\rm eff}}^2, \nonumber
\end{align}
where we have indicated typical values for the parameters.\footnote{One might be tempted to lift this mass by raising $\kappa$, however this implies large fine tuning for electroweak symmetry breaking.} This loop correction almost always dominates over the tree-level mass. \Fig{fig:PGmass} compares the the full one-loop calculation to the estimate in \Eq{eq:oneLoopMassEstimate}.

\subsection{Width with R Violation}

\begin{figure}[t]
\begin{centering}
\subfloat[]{\label{fig:PGmixing1} \includegraphics[width=0.45\textwidth]{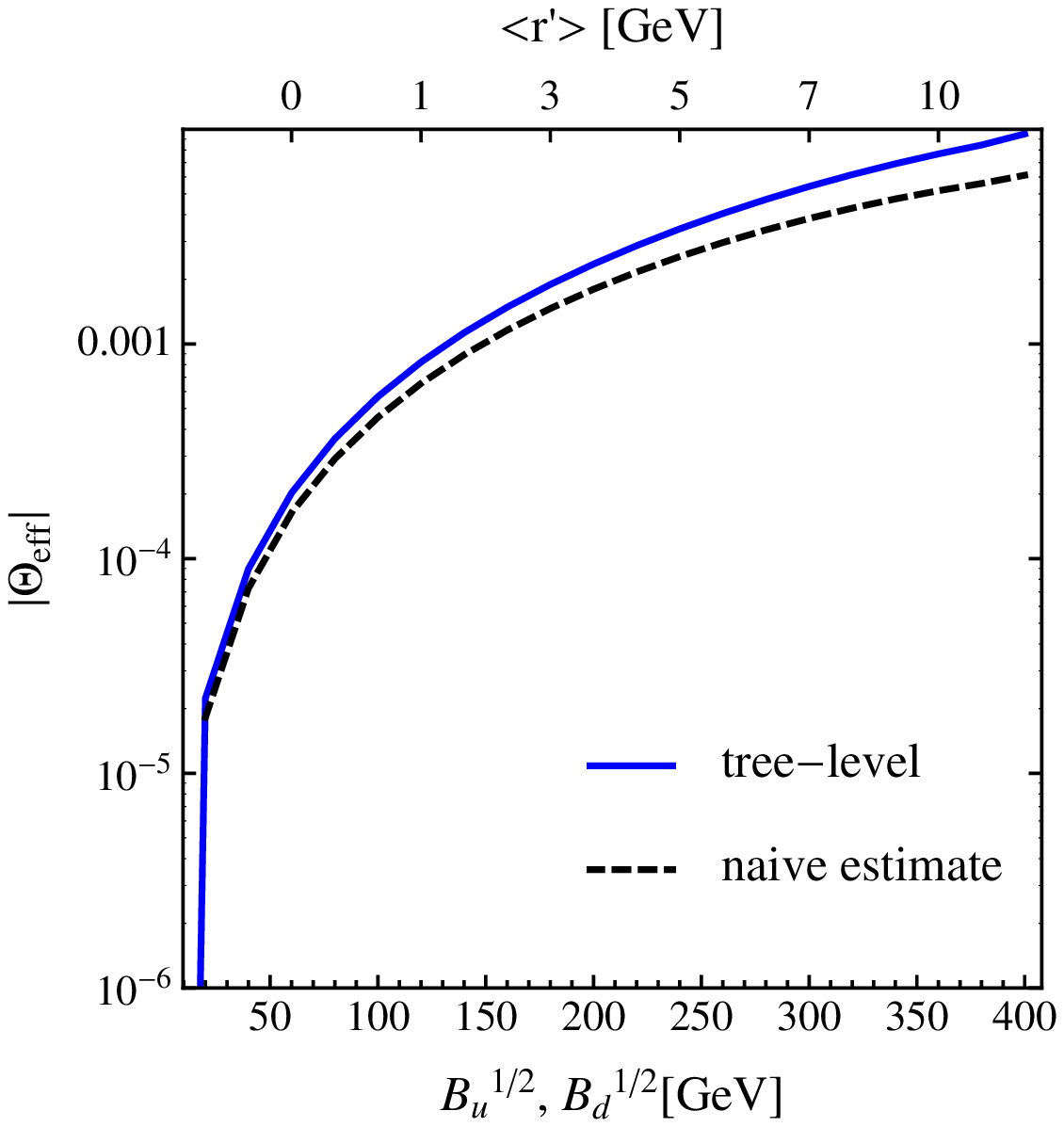}}\hfill
\subfloat[]{\label{fig:PGmixing2} \includegraphics[width=0.45\textwidth]{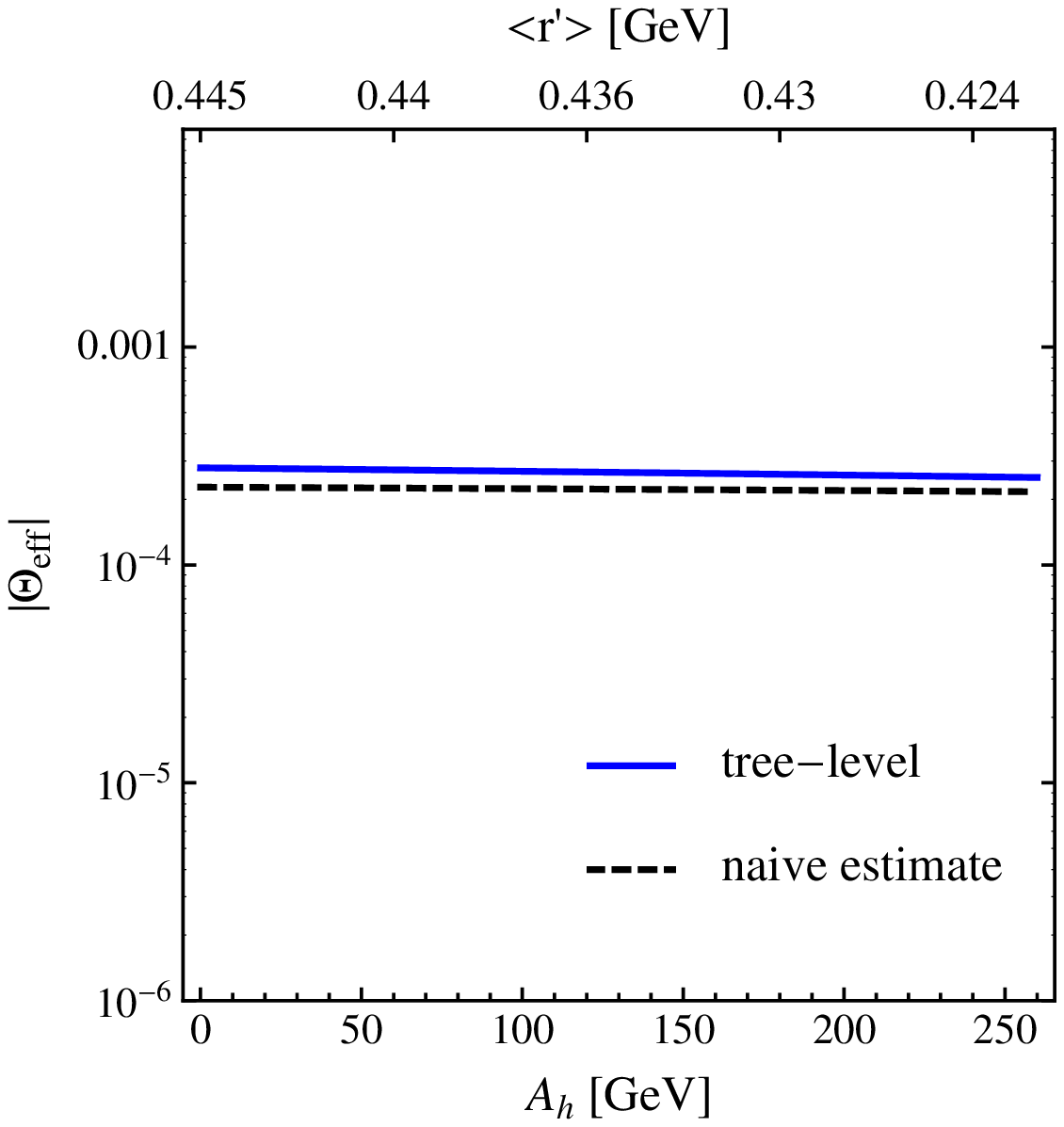}}
\caption{Effective mixing angle of the pseudo-goldstino with gauginos as defined in \Eq{delta}. The left and right plots use the same parameters as \Figs{fig:mass1}{fig:mass2}, respectively.  The blue line represents the exact tree-level result and the dashed line shows the naive estimate according to \Eq{eq:firstordermass}.  As a reference, the value of $\vev{r'}$ is shown, since this controls the mixing angle in  \Eq{eq:firstordermass}.  Notice that in right plot, $\vev{r'}$ is small compared to the left plot and almost constant. According to \Eq{eq:firstordermass}, this leads to a smaller and almost constant mixing angle (and tree-level mass).} 
\label{fig:PGmixing}
\end{centering}
\end{figure}

In the presence of $R$-violating soft masses, the pseudo-goldstino mixes with the bino and neutral wino states.  From \Eq{eq:firstordermass}, we see that this mixing is suppressed, both by gauge couplings and by the small size of the $R$-violating parameter $\vev{r'}$.  The typical mixing angle can be read off from \Eq{eq:firstordermass} by normalizing the state. The full expression is not insightful, however for weak-scale soft parameters we generally obtain
\begin{equation}
\label{eq:mixAngle}
\Theta_{\widetilde{B},\psGld}\sim\Theta_{\widetilde{W},\psGld} \sim \order(10^{-2}-10^{-4}),
\end{equation}
where the range is set by the size of $\vev{r'}$ as illustrated in \Fig{fig:PGmixing}. 

\begin{figure}[t]
\begin{centering}
\includegraphics[scale=0.7]{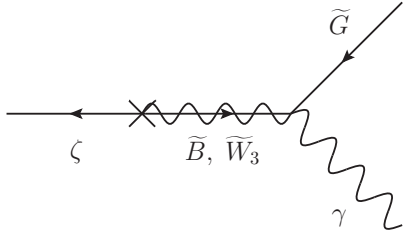}
\caption{Dominant decay mode for the pseudo-goldstino in the $R$-violating case.  This decay occurs through the (small) mixing angle between the pseudo-goldstino and the neutral gauginos. }
\label{fig:psGld_decay}
\end{centering}
\end{figure}

This small mixing with the gauginos induces a coupling of the pseudo-goldstino to the gravitino and photon, permitting the decay $\psGld\rightarrow\gamma + \gravitino$ as shown in \Fig{fig:psGld_decay}.  Since no other final states are kinematically allowed, this is the dominant decay mode of the pseudo-goldstino.

We can calculate the pseudo-goldstino width using the goldstino equivalence theorem.  The longitudinal gravitino $\widetilde{G}_L$ (approximated by the true goldstino in \Eq{chihid}), couples derivatively to the supercurrent:
\begin{equation}
\label{eq:supercurrentInt}
\mathcal{L} =\frac{1}{F}(\partial_\mu \widetilde{G}_L) j^\mu+ \HC
\end{equation} 
The supercurrent contains the coupling of the pseudo-goldstino $\psGld$ to the photon
\begin{equation}
j^\mu\supset-\frac{\Theta_{\mbox{\scriptsize{eff}}}}{2\sqrt{2}}(\sigma^\nu\bar{\sigma}^\rho\sigma^\mu \psGld^\dagger)F_{\nu\rho},
\end{equation}
where the effective ``photino'' mixing angle is determined by the weak mixing angle $\theta_w$,
\begin{equation}\label{delta}
\Theta_{\mbox{\scriptsize{eff}}}=\cos\theta_w\Theta_{\widetilde{B},\psGld}+\sin\theta_w\Theta_{\widetilde{W},\psGld}.
\end{equation}
Using various equations of motion, the interaction term \Eq{eq:supercurrentInt} contains 
\begin{equation}
\label{inter}
\mathcal{L} \supset\frac{\Theta_{\mbox{\scriptsize{eff}}}}{\sqrt{2}F}m_{\psGld}^2(\widetilde{G}_L \sigma^\mu \psGld^\dagger)A_\mu+ \HC
\end{equation}
where $m_{\psGld}$ is the physical mass of the pseudo-goldstino.  The width of the pseudo-goldstino is thus\footnote{Instead of using a derivatively coupled basis for the true goldstino, one could use a non-derivative basis where the goldstino coupling is proportional to the gaugino soft mass $M$.  One might worry that in the non-derivative basis, the decay width would scale as $m_\psGld^3M^2/F^2$ instead of scaling as $m_\psGld^5/F^2$.  However, one can show that a cancellation occurs when proper mixing angles are taken into account, namely $\cos\theta_w\Theta_{\widetilde{B},\psGld}M_1+\sin\theta_w\Theta_{\widetilde{W},\psGld}M_2\equiv m_\psGld\Theta_{\mbox{\scriptsize{eff}}}$, and the two bases give consistent results.}
\be
\label{eq:gamma}
\Gamma(\psGld\rightarrow\gamma + \widetilde{G}_L) = \frac{|\Theta_{\mbox{\scriptsize{eff}}}|^2}{16\pi F^2}m_{\psGld}^5.
\ee
The lifetime of the pseudo-goldstino is
\be
\label{eq:lifetime}
\tau \equiv \frac{1}{\Gamma_\psGld} \simeq  10^{9}~\text{sec} \parfrac{10^{-3}}{\Theta_{\rm eff}}^2 \parfrac{F}{10^{10}~\gev^2}^2 \parfrac{100~\mev}{m_{\psGld}}^5,
\ee
which is generically a cosmological problem, as discussed further in \Sec{sec:Cosmo}.

\subsection{The R-symmetric Case}
\label{sec:RSmass}

Because of the cosmological difficulties in the $R$-violating case, it is worthwhile to consider the possibility that the visible sector $R$-symmetry is not violated by SUSY breaking mediation from the hidden sector.  In this case, only the soft masses $\widetilde{m}^2$ in \Eq{eq:soft} are relevant.  As in the minimal $R$-symmetric SSM \cite{Kribs:2007ac}, we can generate Dirac gaugino masses by introducing chiral superfields $\Phi_i$ in the adjoint representation of the SM gauge groups \cite{Fox:2002bu}: 
\be
\label{eq:diracmasses}
\int d^2\theta\frac{\theta_{\alpha} D'}{\Lambda} W_i^{\alpha} \Phi_i,
\ee
where $\theta_{\alpha} D'$ is a $D$-type spurion with $R$-charge $1$, and the index $i$ runs over the SM gauge groups.  The fermionic components of $\Phi_i$ marry the SSM gauginos with a Dirac mass term proportional to $D'/\Lambda$.\footnote{A similar mechanism could generate a Dirac mass for the pseudo-goldstino, a possibility we will not pursue.}

As touched on in \Sec{subsec:HidSusy}, an exact $R$-symmetry in the visible sector implies an exactly massless state, which in the sequestered limit corresponds to the goldstino of the visible sector.  Of course, there is an irreducible contribution to $R$-violation from SUGRA, since canceling the cosmological constant by hand explicitly violates any $R$-symmetry.   In \Ref{Cheung:2010mc}, it was argued that if two sequestered sectors independently break SUSY and couple solely through SUGRA, then one linear combination of the goldistini is eaten by the gravitino, while the orthogonal combination obtains a mass $2\mgrav$.  However, in the present case, the SUSY breaking sectors are not even approximately sequestered, since the hidden sector is necessary to achieve weak-scale superpartners and evade the supertrace sum rule.

It is straightforward to calculate the SUGRA contribution to the pseudo-goldstino mass (for example, using the methods introduced in \Ref{Cheung:2011jp,Cheung:2011jq}), but a toy model is sufficient to understand the parametric scaling.  Consider a visible sector Lagrangian with a single chiral multiplet $X$
\be
\label{eq:toyLag}
\kahler = \hc{X}X -\frac{1}{2 \Lambda^2} \left(\hc{X}X\right)^2, \qquad \sPot = \mu_{x}X . 
\ee
In the absence of SUGRA, the higher-dimensional K\"ahler term stabilizes the sgoldstino $x$ at 0 with a mass 
\be
(m^{\vis}_x)^2 = \frac{\mu^2_x}{\Lambda^2},
\ee
where the ``\vis'' superscript indicates that this is the contribution from the visible sector alone.  For small field vevs, the K\"ahler term also implies a mass term for the pseudo-goldstino\footnote{For larger field vevs, we would have to account for the change in kinetic normalization of the $X$ multiplet.  We are implicitly assuming $\vev{x} \ll \Lambda$.}
\be
m_{\zeta} = -2 \frac{\mu_x \vev{x}^\dagger}{\Lambda^2}.
\ee
where the factor of $2$ is a Majorana symmetry factor.  The leading SUGRA effect is to generate a tadpole term for $x$ proportional to $m_{3/2}$,\footnote{In the conformal compensator formalism, these terms arise from $\sPot \rightarrow \Phi^2 (\mu_x X)$ where $\Phi \simeq 1 + \theta^2 m_{3/2}$.  For large field vevs, there are additional contributions to the mass from the K\"ahler potential discussed in \Ref{Craig:2010yf} and detailed in \Ref{Cheung:2011jq}.}  
\be
\label{eq:tadpoleL}
\mathcal{L} = 2 \mgrav \mu_x x  + \HC
\ee
The $x$ scalar is then stabilized away from zero due to this tadpole, giving rise to a pseudo-goldstino mass in agreement with \Ref{Cheung:2010mc},  
\be
\text{Visible Sector Only}: \quad \vev{x}  = - \frac{\mu_x}{(m^{\vis}_x)^2} m_{3/2}= - \frac{\Lambda^2}{\mu_x} m_{3/2}, \qquad m_{\zeta} = 2 m_{3/2}.
\ee

If SUSY breaking is mediated from the hidden sector, this will generate an additional soft mass term for $x$, $(m^{\hid}_x)^2 |x|^2$.\footnote{We are considering the limit $\vev{F_\vis} \ll \vev{F_\hid}$ so we can ignore modifications to the fermion mass matrix from the hidden sector goldstino.}  The new scalar mass is:
\be
\label{eq:egSoftTerm}
(m^{\tot}_x)^2 =  (m^{\vis}_x)^2 + (m^{\hid}_x)^2.
\ee
Thus, in the presence of the hidden sector, $x$ is stabilized at a different (typically smaller) field value:
\be
\text{Visible \& Hidden Sectors}: \quad \vev{x} = - \frac{\mu_x}{(m^{\tot}_x)^2} m_{3/2}, \qquad m_{\zeta} = 2 \frac{(m^{\vis}_x)^2}{(m^{\tot}_x)^2} m_{3/2}.
\ee
The degree to which the soft mass from the hidden sector dominates the visible sector K\"ahler mass is the degree to which the pseudo-goldstino is lighter than $2m_{3/2}$.  This feature of the toy model is shared by the model in \Sec{sec:model} albeit with complications coming from the fact the pseudo-goldstino is a linear combination of the visible sector fermions and the sgoldstino is generically not a mass eigenstate.  Numerically, the pseudo-goldstino ends up being a few orders of magnitude lighter than the gravitino.

\section{Cosmological Constraints}
\label{sec:Cosmo}

It is well known that long-lived particles with masses above $1~\kev$ can be cosmologically dangerous if they are produced in the early universe; this is the usual gravitino problem \cite{Pagels:1981ke}.   This implies significant cosmological constraints on the pseudo-goldstino in the $R$-violating case, since it has a mass in the range $\order(10~\mev-1~\gev)$ and a lifetime that is typically longer than a second.  In contrast, $R$-symmetric mediation yields a pseudo-goldstino lighter than the gravitino, which can be as light as a few eV.  We discuss the cosmological implications of each scenario in turn.

In the $R$-violating case, stringent bounds apply to the pseudo-goldstino because it is generically a long-lived hot relic.  From \Eq{eq:lifetime}, the pseudo-goldstino has a lifetime which is typically much longer than the time at which Big Bang Nucleosynthesis (BBN) begins ($t_{\rm BBN} \approx 1~\text{sec}$).  In principle, a long lifetime is not constrained as long as the energy density stored in the pseudo-goldstino is much smaller than the radiation energy density at the time of BBN.  However, this is not the case, as shown in \App{app:annihilation}.  The pseudo-goldstino has couplings which are strong enough to allow it to be in thermal equilibrium with the SSM when the temperature of the universe is above the weak scale.  But the couplings of the pseudo-goldstino are sufficiently small that the pseudo-goldstino freezes out while it is still relativistic, leading to a large number density $n_{\psGld} \propto T^3$ and a correspondingly large energy density $\rho_{\psGld} \propto m_{\psGld} T^3$ which is grossly at odds with BBN for masses in the range $\order(10~\mev-1~\gev)$.

The only way to avoid these BBN constraints is for the pseudo-goldstino to decay more quickly.\footnote{Alternatively, one could try to arrange additional annihilation channels for the pseudo-goldstino such that it becomes a cold relic.}  In fact, for sufficiently low hidden sector breaking and large enough $R$-violation (in the form of $\vev{x}$ and $\vev{r'}$) the lifetime can be short enough to decay before BBN.  With a maximally favorable spectrum with the lowest scale of SUSY breaking, we can achieve
 \be
 \label{eq:lifetime2}
 \tau_{\psGld} \approx 5 \times 10^{-3}~\text{sec} \parfrac{7 \times 10^{-3}}{\Theta_{\rm eff}}^{2} \left(\frac{F}{10^{8} \; \gev^2}\right)^2 \left(\frac{1 \; \gev}{m_{\psGld}}\right)^5,
\ee 
which is cosmological safe (though perhaps unrealistically optimistic).  Note that the decay rate scales as the fifth power of the mass, but the arguments in \Sec{eq:whyLight} preclude a pseudo-goldstino heavier than a few GeV.  Alternatively, it is possible that the universe did not reheat up to the weak scale, in which case our cosmological considerations are not applicable.  

A more favorable cosmology occurs if the mediation is $R$-symmetric.  As discussed in \Sec{sec:RSmass}, the pseudo-goldstino is then much lighter than the gravitino.  For light gravitino masses that evade cosmological constraints ($m_{3/2} \lsim 1~\kev$), the pseudo-goldstino is also cosmologically safe.\footnote{In the case that the gravitino is much heavier ($\order(100 \; \gev)$), the bounds discussed in the $R$-violating case would apply to the pseudo-goldstino.  In particular, $R$-symmetry is no longer a good symmetry since the pseudo-goldstino feels substantial $R$-violation from anomaly mediation \cite{Randall:1998uk,Giudice:1998xp}, pushing its mass above $1~\kev$.}  This is because the pseudo-goldstino never carries an appreciable fraction of the total energy density of the universe, and its contribution at late times is further diluted by the QCD phase transition.  

\section{Collider Phenomenology}
\label{sec:LHC}

With visible sector SUSY breaking, there can be dramatic effects on collider phenomenology from the presence of new light states below the weak scale.  We have seen that there is always a light pseudo-goldstino in the spectrum.  As we will explain in more detail in \Sec{sec:Sgold}, there is also typically a light complex scalar which is related to the sgoldstino and denoted by $\phi$.  

These light pseudo-(s)goldstino states affect the decay widths of SSM particles.  As is evident from the superpotential in \Eq{eq:W}, the only couplings of the pseudo-(s)goldstino to the SSM are through the Higgs sector and the gauge sector. Therefore, the presence of these light states generically alters the decay width of the Higgs boson and the lightest neutralino (since it is a linear combination of fields originating in the Higgs and gauge sectors).  

A detailed discussion of modified Higgs decays appears in \Sec{sec:ModHiggs}.  We will find that the pseudo-sgoldstino generally dominates the Higgs width as $h^0 \rightarrow \phi \phi$ if such a decay is kinematically allowed. This implies that the Higgs is not SM-like in its decays, which is particularly relevant in light of the LHC's rapid march through the entire mass range of the SM Higgs.  The exact final state of the Higgs decay depends on how much $R$-violation is in the low energy Lagrangian.  In \Sec{sec:NeutralinoDecays}, we discuss the lightest observable-sector supersymmetric particle (LOSP), focusing on the case of a neutralino LOSP.  In contrast to typical light-gravitino phenomenology, a neutralino LOSP dominantly decays to the pseudo-goldstino rather than the gravitino, and typically in association with a $Z$.

\subsection{The Pseudo-Sgoldstino}
\label{sec:Sgold}

Spontaneous SUSY breaking in a Wess-Zumino model leads to a sgoldstino, namely, a complex scalar that is massless at tree-level and which is the superpartner of the goldstino.  Its mass is in general lifted by loops within the sector that breaks SUSY.  Thus, if the hidden and visible sectors were completely sequestered, we would expect a light complex scalar that corresponds to the pseudo-sgoldstino direction.

\begin{figure}
\begin{center}
\includegraphics[width=0.45\textwidth]{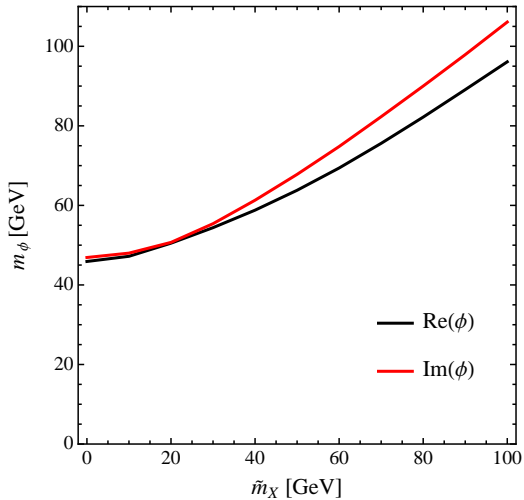}
\end{center}
\caption{An illustration of the pseudo-sgoldstino mass as a function of the soft mass $\widetilde{m}_X$.  We fix $\widetilde{m}_{H_u}^2=\widetilde{m}_{H_d}^2=\widetilde{m}_{R_u}^2=\widetilde{m}_{R_d}^2=(140~\gev)^2$, $\mu_u=300~\gev$, $\mu_d=500~\gev$, $B_u=B_d=(100~\gev)^2$, $\lambda=1$, $\kappa$ is fixed by $m_Z$, and all other soft parameters are set to zero.  Shown are the masses of both the scalar and pseudo-scalar components of $\phi$, which split as $\widetilde{m}_X$ increases.  Since $X$ is a singlet, $\widetilde{m}_X$ is expected to be small in many mediation schemes.   In the $R$-symmetric case (i.e. $B_u = B_d = 0$), the behavior is qualitatively similar except $\Re\, \phi$ and $\Im\,\phi$ are degenerate.} 
\label{fig:sgoldsinomass}
\end{figure}

In the presence of soft masses generated from hidden sector mediation, the pseudo-sgoldstino $\phi$ can get a weak-scale mass.  In practice, though, as long as the $X$ soft mass is small, then $\phi$ will be lighter than the weak scale.  One motivation for a small $X$ soft mass is that if the gravitino is light, then SUSY breaking mediation is most easily achieved via gauge interactions, and therefore it is reasonable to assume that the soft mass for $X$ is small relative to other SSM soft masses since it is a gauge singlet.  With this assumption, the pseudo-sgoldstino is mostly aligned along the $X$ direction and is the lightest scalar in the spectrum.  In \Fig{fig:sgoldsinomass}, we show how the mass of $\phi$ changes as the $X$ soft mass is varied.

A light pseudo-sgoldstino has important consequences for collider phenomenology because it opens new decay modes for the Higgs boson and the LOSP. We discuss this further in the following subsections, currently focusing on the decay modes of the pseudo-sgoldstino itself.

If the soft parameters violate the $R$-symmetry, typically $x$, $r^0_u$, and $r^0_d$ get vevs proportional to a linear combination of $B_u$ and $B_d$.  This implies that all of the neutral scalars ($x,h^0_u,h^0_d,r^0_u,r^0_d$) mix, which in turn gives the light pseudo-sgoldstino decay modes to SM fermions through the SSM Yukawa couplings.  The pseudo-sgoldstino is generically more massive than the $b \bar{b}$ threshold and tends to decay through this channel.  The width is
\be
\label{eq:widthPhiBB}
\Gamma_{\phi\rightarrow b\bar{b}} = \frac{3 }{16 \pi^2}y^2_b |\Theta_{h^0_d,\phi}|^2m_{\phi}\left(1-\frac{4m^2_b}{m^2_{\phi}}\right)^{3/2},
\ee
where $y_b$ is the bottom Yukawa coupling.  This decay is prompt on collider scales for any reasonable value of $\Theta_{h^0_d,\phi}$.

In the case that the mediation is $R$-symmetric, there is an irreducible contribution to $R$-violation from SUGRA effects.  At tree-level in SUGRA, the soft terms
\be
B_u \simeq m_{3/2} \mu_u, \qquad B_d \simeq m_{3/2} \mu_d, \qquad T \simeq 2m_{3/2} \kappa,
\ee
are generated.   After electroweak symmetry breaking, this implies that $\phi$ will have small mixings with the neutral Higgses and can therefore decay to SM fermions.  In particular, the mixing angle $\Theta_{h^0_d,\phi}$ in \Eq{eq:widthPhiBB} scales as $\Theta_{h^0_d,\phi} \simeq \mgrav \mu_d /(\mu^2_d +\ssoftmass{h_d})$, leading to 
\be
\label{eq:widthPhiBBwithR}
\Gamma_{\phi\rightarrow b\bar{b}} & \approx & 10^4 \; {\rm sec}^{-1} \parfrac{y_b}{0.05}^2 \parfrac{\mgrav}{1 \; \kev}^2 \parfrac{300 \; \gev}{\mu_d}^2\parfrac{1}{1 +\ssoftmass{h_d}/\mu^2_d} \parfrac{m_{\phi}}{50 \; \gev}.
\ee
However for the cosmologically preferred region $m_{3/2} \ll \kev$, the $R$-symmetric decay $\phi \rightarrow \psGld \gravitino$ dominates over $\phi \rightarrow b \bar{b}$.  The width of this channel is 
\be
\Gamma_{\phi \rightarrow \psGld \gravitino} = \frac{1}{16 \pi} \left|\Theta_{x,\phi}\Theta_{\widetilde{x},\psGld}\right|^2 \frac{m^5_{\phi}}{F^2} \approx 10^{4} \; {\rm sec}^{-1}\parfrac{\Theta_{\phi,x}\Theta_{\widetilde{x},\psGld}}{1.0}^2 \parfrac{1 \; \kev}{\mgrav}^2 \parfrac{m_{\phi}}{50 \; \gev}^5,
\ee 
where we have neglected mixing with the $R_i$ fields.  Since $\mgrav$ is expected to be lighter than $1~\kev$ to avoid the cosmological gravitino problem, we expect $\phi$ to have an invisible decay in the $R$-symmetric case.

\subsection{Modified Higgs Decays}
\label{sec:ModHiggs}

\begin{figure}
\begin{center}
\subfloat[]{\raisebox{0mm}{{\label{fig:h_2phi}}\includegraphics[scale=0.55]{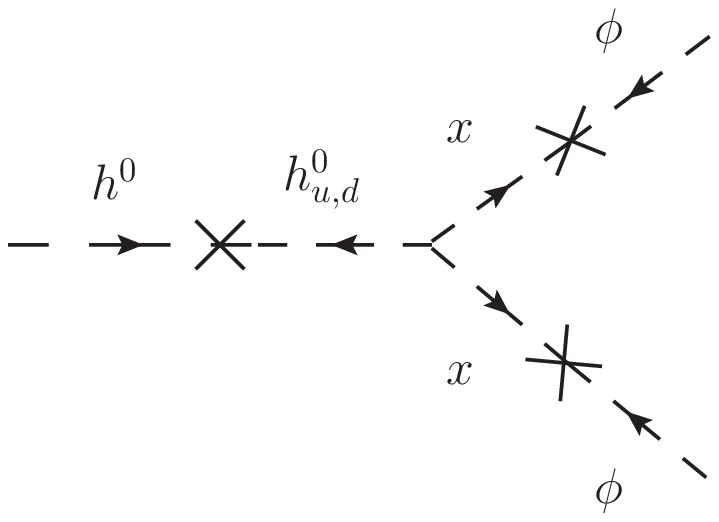}}} $\qquad$
\subfloat[]{\raisebox{5mm}{{\label{fig:phi_bb}}\includegraphics[scale=0.55]{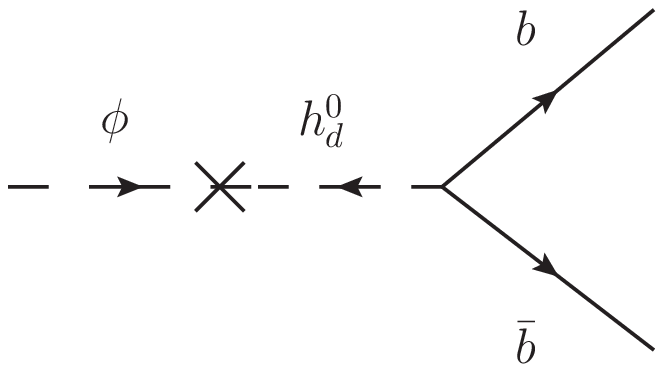}}}  $\qquad$
\subfloat[]{{\label{fig:phi_in}}\includegraphics[scale=0.55]{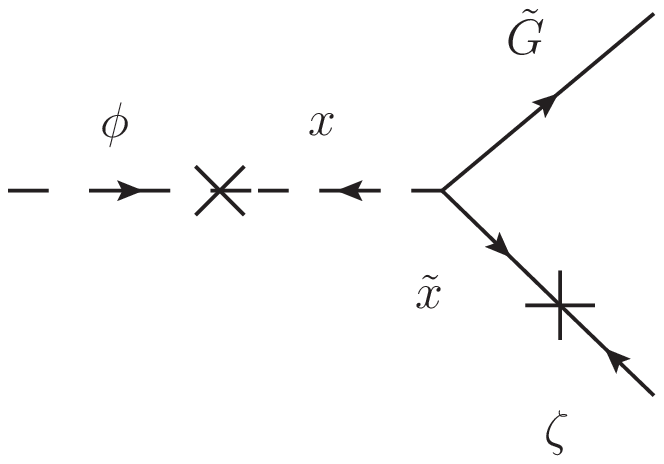}}
\end{center}
\caption{Representative diagrams that modify the Higgs width.  On the left, we show the dominant decay mode of the Higgs when the pseudo-sgoldstino is light:  $h^0 \rightarrow \phi \phi$. The pseudo-goldstino decays dominantly as $\phi \rightarrow b \bar{b}$ in the $R$-violating case (middle), while it decays as $\phi \rightarrow \psGld \gravitino$ in the $R$-symmetric case with a sub-keV gravitino (right).}
\label{fig:modHiggsDiagrams}
\end{figure}

Perhaps the most interesting prediction of visible SUSY breaking is that the Higgs boson will cascade decay through two pseudo-sgoldstinos if it is kinematically allowed.  This statement is independent of the mediation mechanism because the interaction between the Higgs boson and the pseudo-sgoldstino arises from the ($R$-symmetric) $F$-term potentials from $H_u$ and $H_d$.  The decay $h^0 \rightarrow \phi \phi$ is reminiscent of certain regions of NMSSM parameter space \cite{Chang:2008cw}.

The final state of the Higgs cascade decay depends on the amount of $R$-violation in the visible sector.  The decay $h^0 \rightarrow \phi \phi \rightarrow b\bar{b} b\bar{b}$ is expected in the $R$-violating case, and the invisible final state $h^0 \rightarrow \phi \phi \rightarrow \psGld \gravitino \psGld \gravitino$ is expected in the $R$-symmetric case with a light gravitino, as shown in \Fig{fig:modHiggsDiagrams}.  Such decays are particularly interesting now, since LHC data is rapidly ruling out large swaths of SM-like Higgs decays \cite{ATLAS-CONF-2011-157,CMS-PAS-HIG-11-023}.  Exotic final states might delay the observation of the Higgs, especially in the low mass region, because they suppress the branching ratios to more easily detected channels such as $\gamma \gamma$ and $WW^{(*)}$.  That said, the discovery prospects for the $4b$ \cite{Cheung:2007sva,Carena:2007jk} and invisible \cite{Eboli:2000ze,Godbole:2003it, Davoudiasl:2004aj,ATL-PHYS-PUB-2009-061,CERN-THESIS-2010-152} final states are promising.

\begin{figure}[t]
\begin{centering}
\subfloat[]{\label{fig:brHiggsSweep} \includegraphics[width=0.45\textwidth]{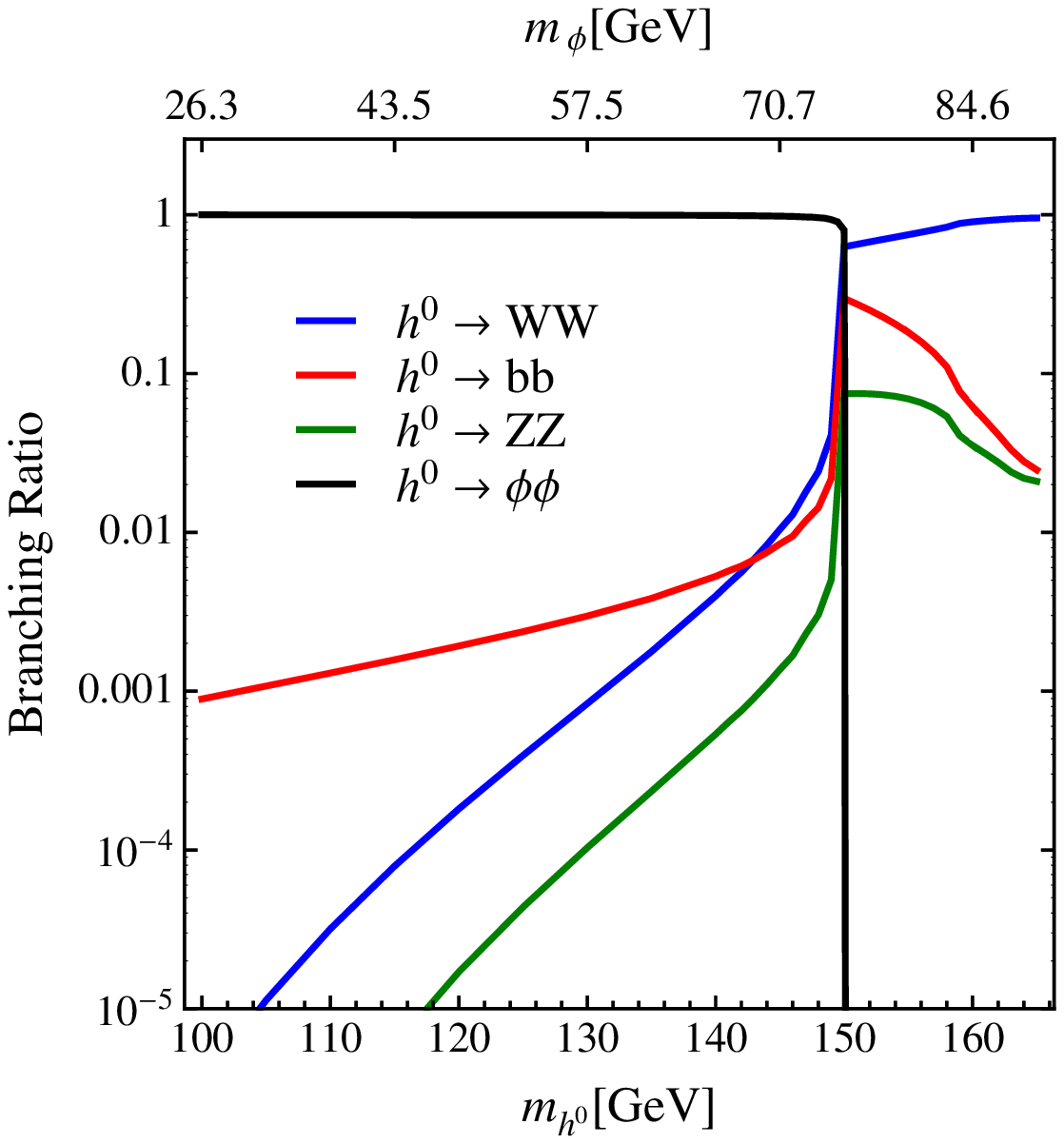}}\hfill
\subfloat[]{\includegraphics[width=0.41\textwidth]{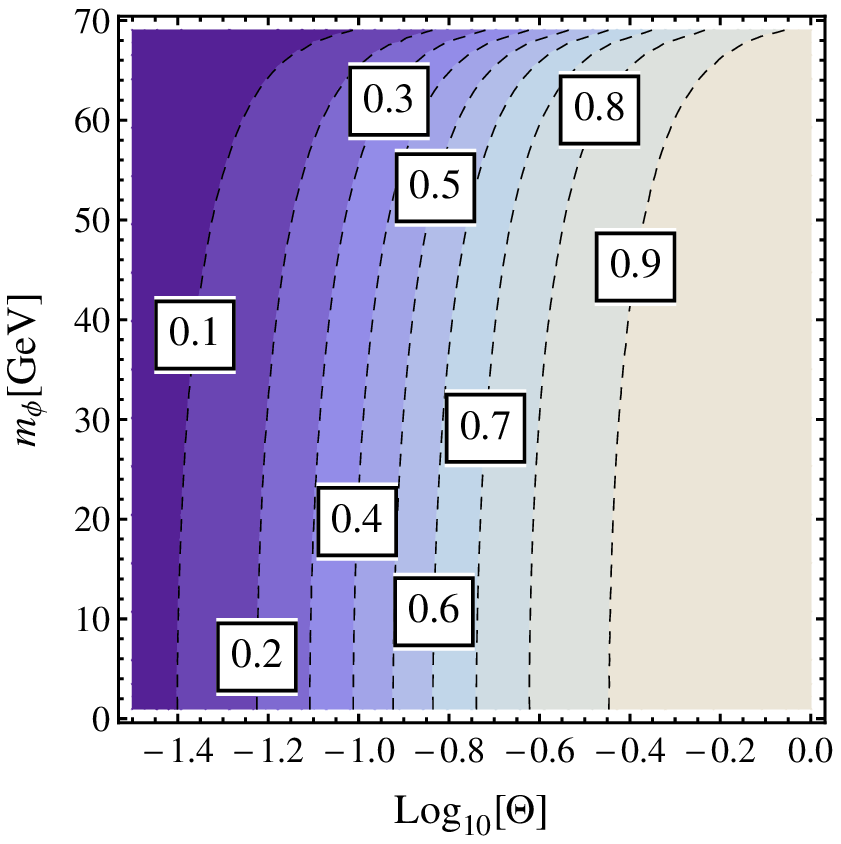}}\hfill
\caption{Left:  the Higgs branching ratios as a function of its mass in the $R$-symmetric case. We fix $\widetilde{m}_{H_u}^2=\widetilde{m}_{H_d}^2=(140~\gev)^2$, $\widetilde{m}_{R_u}^2=\widetilde{m}_{R_d}^2=(150~\gev)^2$, $\mu_u=150~\gev$, $\lambda=1$, $\kappa$ is fixed by $m_Z$, and all other soft parameters are set to zero. We have traded $\mu_d$ for $m_{h^0}$, and the corresponding value of $m_\phi$ is indicated for reference.  When kinematically allowed, the decay $h^0\rightarrow \phi\phi$ is dominant.  Right:  the branching ratio for $h^0\rightarrow \phi\phi$ as a function of $m_\phi$ and the mixing angle $\Theta$ according to \Eq{eq:higgs2phiphi}.  Here, we have fixed $m_{h^0}=140~\gev$, $\lambda=1$, and assumed SM decay widths to $b\bar{b}$, $ZZ^{*}$, and $WW^{*}$.}
\label{fig:higgs_decay}
\end{centering}
\end{figure}

The interaction leading to a modified Higgs decay is
\begin{equation}
\mathcal{L}_{\rm int}\supset\lambda^2|x|^2\left(|h_u^0|^2+|h_d^0|^2\right),
\end{equation}
which yields a decay width
\begin{equation}
\label{eq:higgs2phiphi}
\Gamma(h^0\rightarrow \phi\phi)=\frac{\lambda^4 |\Theta|^2}{16\pi}\frac{v_{\rm EW}^2}{m_{h^0}}\left(1-\frac{4m_{\phi}^2}{m_{h^0}^2}\right)^{1/2},
\end{equation}
where $v_{\rm EW}=\sqrt{\langle h_u^0\rangle^2+\langle h_d^0\rangle^2}$ and  $\Theta \simeq \Theta_{h^0_{u,d},h^0}\Theta^2_{x,\phi}$ .  We find that $\Theta$ is typically of $\order(0.3)$ or greater. We estimate the width as
\begin{equation}
\Gamma(h^0\rightarrow \phi\phi)\approx 10^{-1}~\gev \parfrac{\lambda}{1.0}^2 \left(\frac{\Theta}{0.3}\right)^2\left(\frac{150\mbox{~GeV}}{m_{h^0}}\right),
\end{equation}
which can easily dominate over SM decay channels.  In \Fig{fig:higgs_decay}, we illustrate the branching ratios of $h \rightarrow\{ \phi\phi, \; b\bar{b}, \; WW^{(*)},\;ZZ^{(*)} \}$ for a representative sweep of parameter space.  As advertised, if $h\rightarrow \phi \phi$ is kinematically allowed, then it dominates the width.\footnote{Depending on the region of parameter space, this can even be true above the $WW$ threshold.} We note that the physical Higgs mass can be significantly larger than $m_{Z}$ because $\lambda$ contributes to the Higgs quartic coupling.  This fact is reflected in \Fig{fig:brHiggsSweep}.

In addition to the SM-like Higgs, the enlarged Higgs sector can give rise to a rich phenomenology.  While a full study is beyond the scope of this paper, we wish to highlight some interesting features.  In the $R$-symmetric case, heavier scalars in the Higgs sector are
neatly separated between $R_{u,d}$-like states and $H_{u,d}$-like states because the mixing is proportional to $m_{3/2}$.  Searching for $R_{u,d}$-like states would help to distinguish our scenario from the NMSSM. While single production of $R_{u,d}$ is heavily suppressed by $\langle r_{u,d}^0 \rangle/v_{\rm EW} \ll1$, $R_{u,d}$ can be produced in the decays of heavier states, as well as through electroweak pair production \cite{Choi:2010ab}.  The neutral $R_{u,d}$-like scalars typically decay to $h^0 \phi$ or $\chi^0\psGld$, where $\chi^0$ is the lightest neutralino.\footnote{Subsequent decays of the $\chi^0$ are discussed in \Sec{sec:NeutralinoDecays}.}  The charged $R_{u,d}$ states typically decay to $\chi^\pm \psGld$ or $W^\pm R^0$, where $R^0$ is the lightest $R_{u,d}$-like neutral state, so one expects the $R_{u,d}$-like decays to be invisible or semi-invisible.

Among the $H_{u,d}$-like states, the heavy CP-even and CP-odd Higgs-like states $H^0$ and $A^0$ dominantly decay to $t\bar{t}$ for the same parameter sweep as \Fig{fig:brHiggsSweep}.  Since the Higgs decays invisibly but the heavy Higgs state is visible, the heavy Higgs could be a ``Higgs impostor'', although with altered branching ratios with respect to a SM Higgs boson of the same mass \cite{Bellazzini:2009xt,DeRujula:2010ys,Fox:2011qc}.  There do exist regions of parameter space where the heavy Higgs-like states are below the $t\bar{t}$ threshold, in which case they dominantly decay to an $R_{u,d}$-like scalar and $h^0$ if kinematically allowed, leading to an invisible or semi-invisible decay of the heavy Higgs. The charged $H_{u,d}$-like states dominantly decay to $t\bar{b}$ or $\bar{t}b$ for the same parameter sweep as \Fig{fig:brHiggsSweep}.

\subsection{Modified Neutralino Decays}
\label{sec:NeutralinoDecays}

\begin{figure}[H,t]
\begin{centering}
\subfloat[]{\label{fig:chi_phi-psGld_decay}\includegraphics[scale = 0.55]{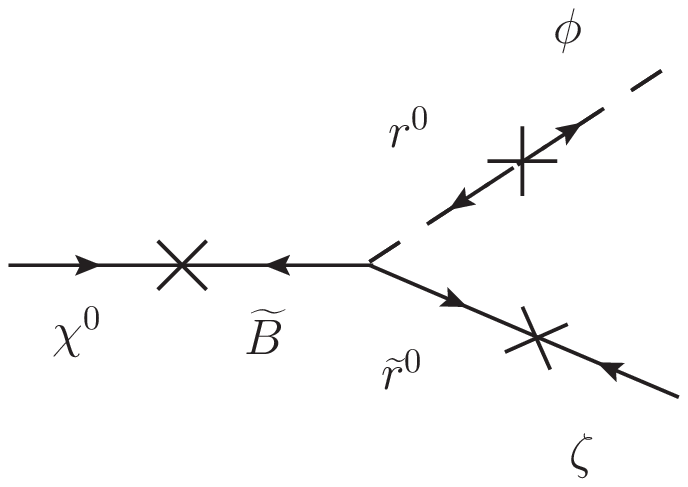}} $\qquad$
\subfloat[]{\label{fig:chi_higgs-psGld_decay}\includegraphics[scale = 0.55]{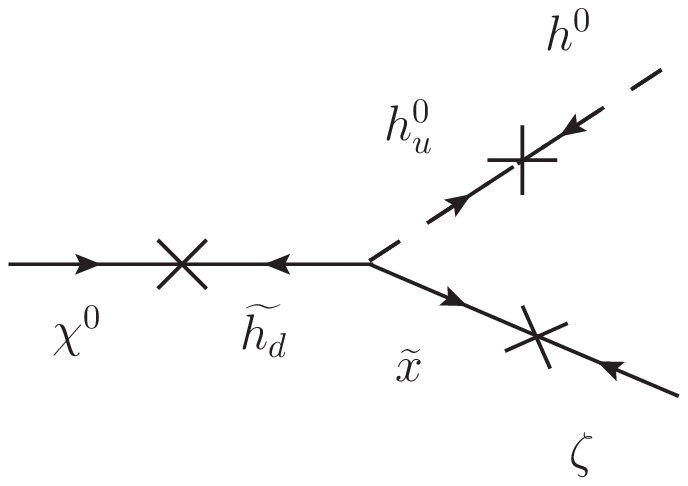}} $\qquad$
\subfloat[]{\label{fig:chi_z-psGld_decay}\includegraphics[scale = 0.55]{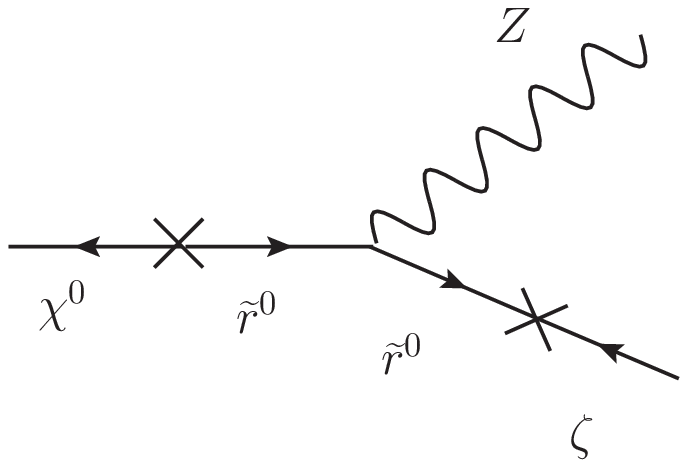}}
\caption{Representative diagrams contributing to the decays of a neutralino LOSP:  $\chi^0 \rightarrow \phi \psGld$ (left), $\chi^0 \rightarrow h^0 \psGld$ (middle), $\chi^0 \rightarrow Z \psGld$ (right).  The relative size of the decay widths is sensitive to the scale of $R$ violation, with $\chi^0 \rightarrow \phi \psGld$ becoming suppressed in the $R$-symmetric limit.  The full set of diagrams appear in \Fig{fig:FullNdecays}.}
\label{fig:neutralinoDecays}
\end{centering}
\end{figure}

As in usual SUSY theories, pair- and associated-production of superpartners result in cascade decays that terminate in two LOSPs.   This follows from the $R$-charge assignments in \Tab{tab1} and the fact that $R$-parity is conserved regardless of whether the $R$-symmetry is broken.  Thus, it is important to identify the decay modes of the LOSP, since this decay will appear in every cascade decay.  For simplicity, we will assume that the LOSP is a neutralino, though other LOSP possibilities can also result in modified phenomenology.

For simplicity, we will ignore decays to gravitinos because are suppressed by the hidden SUSY breaking scale.  We will also ignore decays to photons, since they arise only from higher dimensional operators (since the neutralino is neutral).\footnote{The decay $\chi^0 \rightarrow \gamma \psGld$ can occur from mixing between the visible and hidden sector goldstinos, but this is suppressed by $\vev{F_\vis}/\vev{F_\hid}$.}  The dominant diagrams contributing to these modes for a neutralino LOSP are shown in \Fig{fig:neutralinoDecays}:
\be
\label{eq:RviolatingDecays}
\chi^0 \rightarrow \{ \phi \psGld,  \; h^0 \psGld, \; Z \psGld \}.
\ee
The presence of these decay modes are independent of the $R$-symmetry properties of mediation, however the resulting widths are not.  The $R$-symmetry forbids mixing between the gauginos and the pseudo-goldstino and also forbids mixing between the pseudo-sgoldstino and $h^0_u$/$h^0_d$.  This effect suppresses $\chi^0 \rightarrow \phi \psGld$ in the $R$-symmetric case. 

\begin{figure}
\begin{centering}
\subfloat[]{\includegraphics[width=0.45\textwidth]{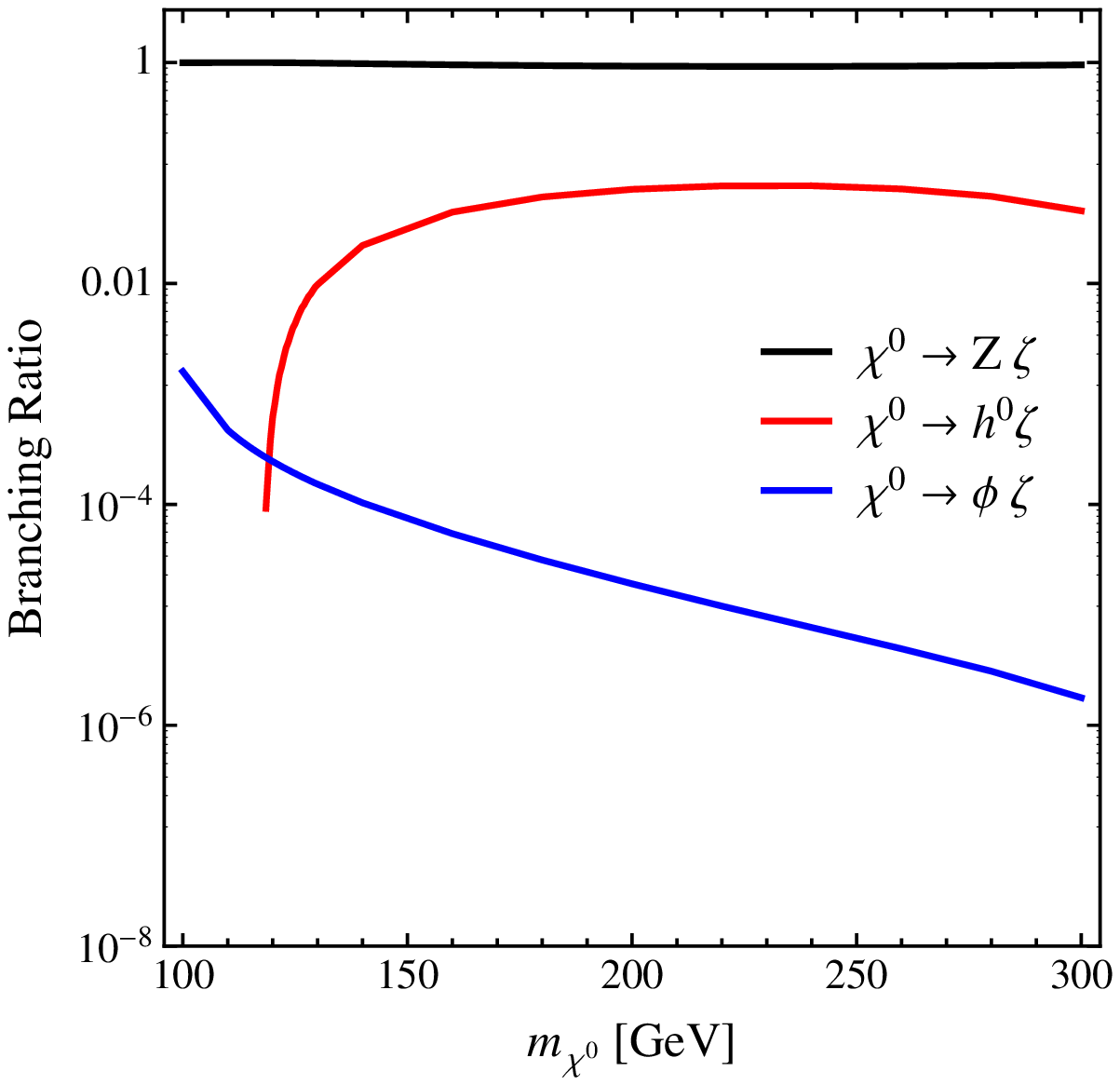}}\hfill
\subfloat[]{\includegraphics[width=0.45\textwidth]{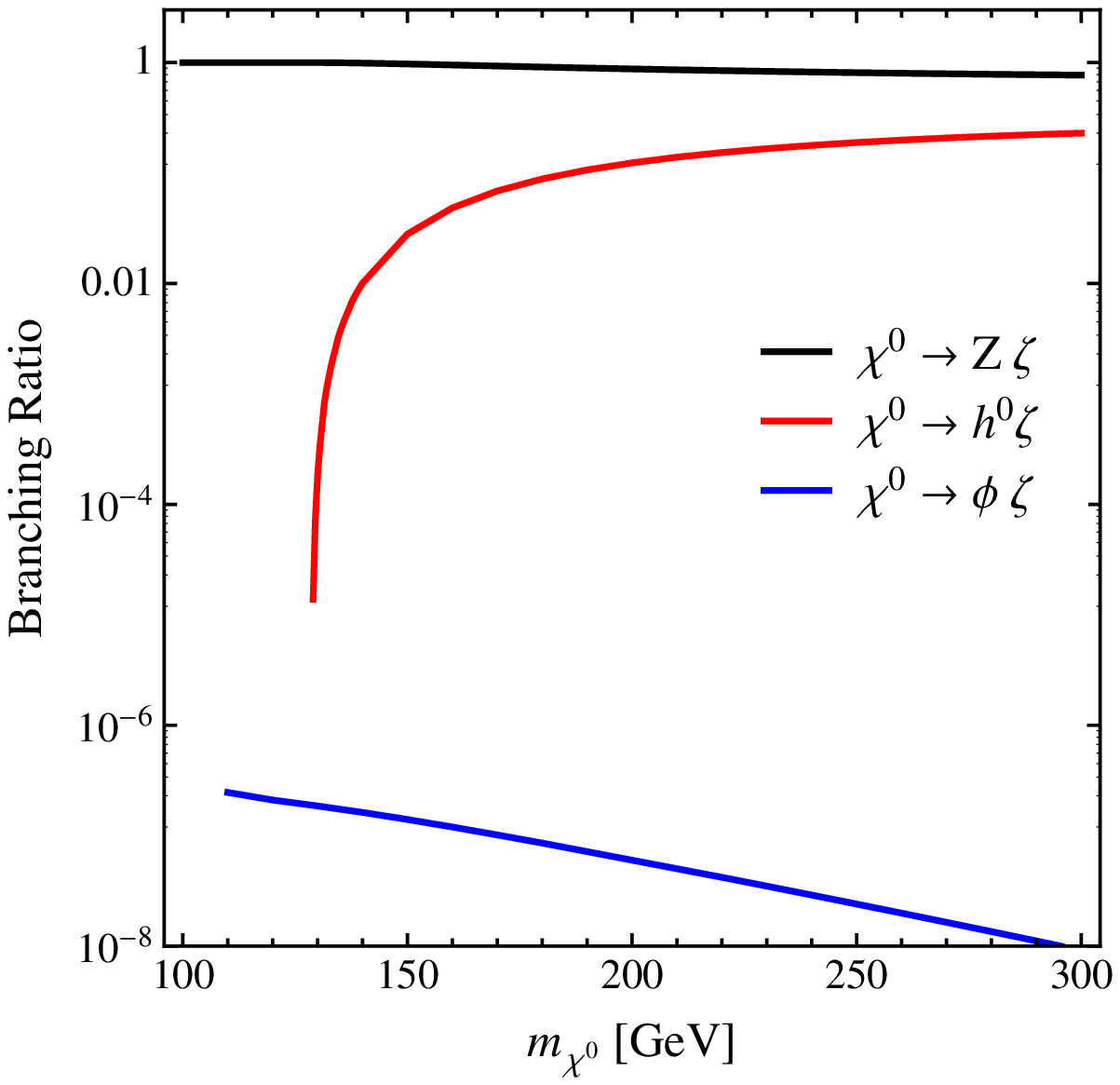}}\hfill
\caption{Left: branching ratios for a neutralino LOSP as a function of its mass $m_{\chi^0}$ in the $R$-violating case.  We fix $\widetilde{m}_{h_u}^2=\widetilde{m}_{h_d}^2=\widetilde{m}_{r_u}^2=\widetilde{m}_{r_d}^2=(140~\gev)^2$, $\mu_u=500~\gev$, $B_u=B_d=(70~\gev)^2$, $M_1=M_2=400~\gev$, $\lambda=1$, $\kappa$ is fixed by $m_Z$, and all other soft parameters are set to zero. We have traded $\mu_d$ for $m_{\chi}$.  Right: branching ratios for a neutralino LOSP in the $R$-symmetric case.  We use the same parameters as in the left figure, except we set the Majorana gaugino masses and $B$-terms equal to zero, and we set the Dirac mass for gauginos $m_D=1\;\tev$.  The dominant decay mode for the neutralino LOSP is $\chi^0 \rightarrow Z \psGld$ over much of parameter space.} 
\label{fig:neutralino_decay}
\end{centering}
\end{figure}

The explicit formulae for the decay widths are given in \App{app:Neutralinos}. We find that in most of the parameter space, $\chi^0 \rightarrow Z \psGld$ is the dominant channel for either mediation scheme, as illustrated in \Fig{fig:neutralino_decay}.  Moreover, this result is largely independent of the higgsino vs.\ gaugino fractions of the LOSP.  One can understand this by examining \Fig{fig:neutralinoDecays} and noting that the typical mixing angles in the $\chi^0 \rightarrow Z \psGld$ diagram are $\order(1)$.  In contrast, the diagrams contributing to $\chi^0 \rightarrow \phi \psGld$ have at least one suppressed mixing, and the decay to Higgs bosons is phase space suppressed. 

Our LOSP decay is similar to a wino-like decay in ordinary gauge mediation, $\widetilde{W}_3 \rightarrow Z + \gravitino$.  However, a distinctive feature is that $Z+\psGld$ dominates even if the mass of the LOSP is comparable to $m_{Z}$.  In ordinary gauge mediation, such a decay is phase space suppressed and $\widetilde{W}_3 \rightarrow \gamma + \gravitino$ becomes dominant. Therefore, observing this decay without an accompanying $\gamma + \gravitino$ channel could provide evidence for the pseudo-goldstino if the LOSP is not too much heavier than $m_{Z}$.\footnote{An additional distinguishing characteristic is that, depending on the mass of the gravitino, neutralino decays in gauge mediation can be displaced whereas decays to the pseudo-goldstino are prompt.}   Similar modified LOSP decays were studied in \Ref{Thaler:2011me}, and a recent study of neutralino LOSP decays in gauge mediation can be found in \Ref{Ruderman:2011vv}.

\section{Conclusion}
\label{sec:conclude}

The possibility that SUSY could be broken in multiple sectors challenges the standard lore concerning the SSM sparticle spectrum.  In particular, the SSM can feel SUSY breaking at tree-level without being constrained by the supertrace sum rule.  The immediate consequence of tree-level SUSY breaking in the SSM is the presence of a light pseudo-goldstino state which mixes with SSM gauginos and higgsinos. 

In this paper, we have studied the simplest extension of the SSM that affords tree-level SUSY breaking.  We expect that many of the conclusions of this paper hold in more generic visible sector SUSY breaking models, since the pseudo-goldstino mass and couplings are largely determined by symmetries.  Phenomenologically, the most important symmetry to understand is a $U(1)_R$ symmetry, and we have argued that the properties of the pseudo-goldstino are sensitive to whether the $R$ symmetry is preserved when hidden sector SUSY breaking is mediated to the SSM. 

In the usual case of $R$-violating soft parameters, the pseudo-goldstino mass is typically one loop factor suppressed relative to the weak scale, and the pseudo-goldstino inherits modest couplings to SSM fields through mixing with the gauginos and higgsinos.  The cosmological constraints on such a state are severe, since a pseudo-goldstino in thermal equilibrium at early times implies overclosure at late times.  In this way, the common assertion that SUSY cannot be broken at tree-level in the SSM still holds, but the reason is pseudo-goldstino cosmology rather than sum rules.  That said, this scenario can be phenomenologically viable if the reheat temperature is $\order(\gev)$ such that the pseudo-goldstino is never in thermal equilibrium.  Also, there are small corners of parameter space where the pseudo-goldstino decays before BBN.
 
Conversely, if the mediation respects the visible sector $R$-symmetry, then the mass of the pseudo-goldstino is protected.  The $R$-violating effects come only from SUGRA, and the pseudo-goldstino mass is proportional to (but generically smaller than) the gravitino mass.  The same region of parameter space that solves the gravitino problem also prevents cosmological overproduction of the pseudo-goldstino.

The distinguishing collider signatures of the simplest visible sector SUSY breaking scenario involve modified Higgs and neutralino decays.  Generically, there exists a light pseudo-sgoldstino $\phi$ that dominates the Higgs width through $h^0 \rightarrow \phi \phi$.  If the mediation is $R$-violating, then this state has mixing with $h_u$ and $h_d$, and the four-body final state $h^0 \rightarrow \phi \phi \rightarrow b\bar{b} b\bar{b}$ is the dominant decay mode.  This is similar to the Higgs phenomenology in some regions of the NMSSM. On the other, if the mediation is $R$-symmetric, then the Higgs boson dominantly decays invisibly to $\psGld \gravitino \psGld \gravitino$, a possibility that is currently being tested at the LHC.  Since the invisible final state involves the gravitino and the pseudo-goldstino, the Higgs sector becomes an interesting probe of spontaneous SUSY breaking dynamics.

\acknowledgments This work was supported by the U.S. Department of Energy under cooperative research agreement DE-FG02-05ER-41360.  J.T. is supported by the DOE under the Early Career research program DE-FG02-11ER-41741. D.B. is partly supported by Istituto Nazionale di Fisica Nucleare (INFN) through a ``Bruno Rossi'' fellowship.

\appendix

\section{One-Loop Pseudo-Goldstino Mass}
\label{sec:oneLoop}

In this appendix, we calculate the pseudo-goldstino mass at one loop, which gives large corrections to the tree-level mass in \Eq{eq:mtree}.  In the mass eigenstate basis, the tree-level neutralino Lagrangian is:
\begin{equation}
\mathcal{L}_\chi=i \overline{{\bs \chi}^0} \bar{\sigma}^\mu\partial_\mu {\bs \chi}^0-\left(\frac{1}{2}({\bs \chi}^0)^T \mathcal{M}_D {\bs \chi}^0+\HC\right)+\mathcal{L}_{\rm int}({\bs \chi}^0, \ldots),
\end{equation}
where $\mathcal{M}_D$ is the diagonal mass matrix.  The one-loop correction to the quadratic Lagrangian can be written as:
\begin{equation}\label{eq:Ln1loop}
\delta\mathcal{L}_{\chi}^{(2)}=i \overline{{\bs \chi}^0} \widehat{\Xi}\bar{\sigma}^\mu\partial_\mu {\bs \chi}^0 -\left(\frac{1}{2}({\bs \chi}^0)^T \widehat{\Omega}{\bs \chi}^0+\HC \right),
\end{equation}
where $\widehat{\Xi}$ and $\widehat{\Omega}$ are properly renormalized self-energy functions.  Using \Eq{eq:Ln1loop}, the one-loop corrected pseudo-goldstino mass is
\begin{equation}\label{eq:mass1}
m_\psGld=(1-\widehat{\Xi}_{\psGld,\psGld})m^{\rm tree}_\psGld+\widehat{\Omega}_{\psGld,\psGld}.
\end{equation}
The tree-level mass $m_\psGld^{\rm tree}$  in \Eq{eq:mtree} already captures the leading contribution from gauge interactions (remember that at tree-level the pseudo-goldstino can get mass \emph{only} through gauge interactions), so at leading order we can ignore corrections coming from $\widehat{\Xi}_{\psGld,\psGld}$.  On the other hand, $\widehat{\Omega}_{\psGld,\psGld}$ is necessary to capture the leading contribution in $\lambda$, such that 
\begin{equation}\label{eq:mass2}
m_\psGld\simeq m^{\rm tree}_\psGld+\widehat{\Omega}_{\psGld,\psGld}.
\end{equation}

\begin{figure}[t]
\begin{center}
\subfloat[]{\includegraphics[scale=0.7]{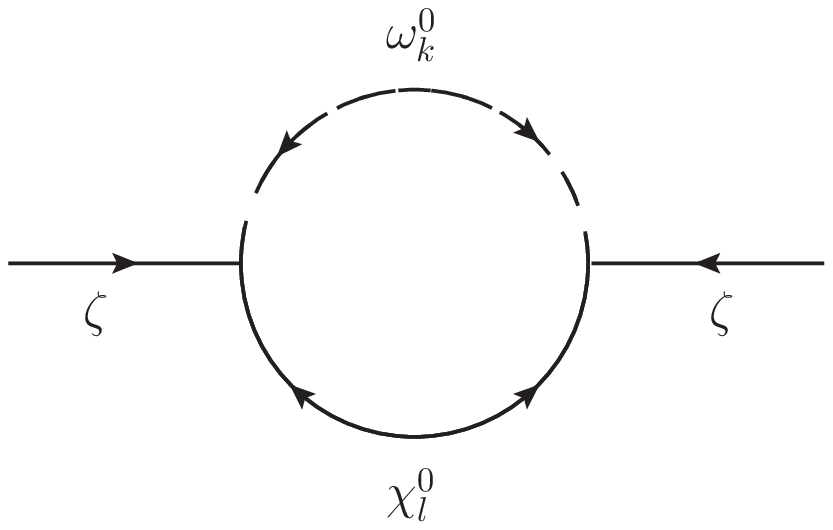}} $\qquad$
\subfloat[]{\raisebox{0.43cm}{\includegraphics[scale=0.7]{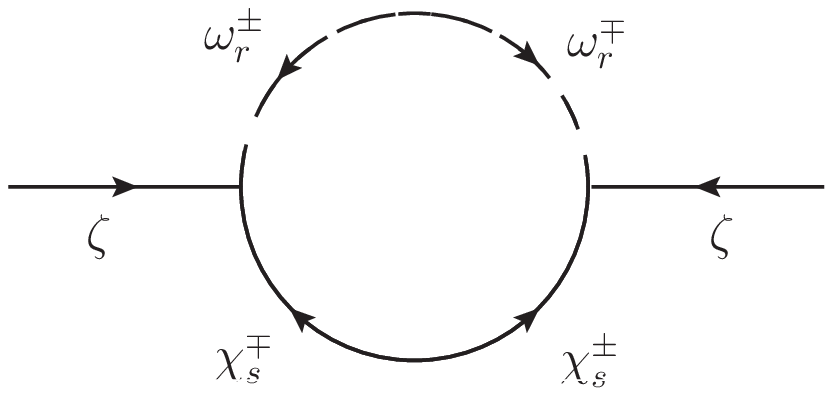}}}
\caption{One-loop self-energy diagrams contributing to the pseudo-goldstino mass.}
\label{fig:feyn1}
\end{center}
\end{figure}

The $\lambda XH_uH_d$ term in \Eq{eq:W} contains the following interactions:
\begin{equation}
\label{eq:Lint}
\mathcal{L}\supset\lambda
\left(-\mathcal{Z}^{k\ell}\omega^0_k \chi^0_\ell+\mathcal{Y}^{rs}\omega^+_r\chi^-_s
+\mathcal{J}^{rs}\omega^-_r \chi^+_s\right)\psGld+\HC,
\end{equation}
where $\omega^0_k$ and $\omega^\pm_r$ are the neutral and charged scalar mass eigenstates (including Goldstone bosons), and $\chi^0_\ell$ and 
$\chi^\pm_s$ are the neutralino and chargino mass eigenstates. The matrices $\mathcal{Z}$, $\mathcal{Y}$, and $\mathcal{J}$ encode the appropriate mixing angles between gauge and mass eigenstates. The one-loop Feynman diagrams generated by \Eq{eq:Lint} and contributing to the bare self-energy $\Omega_{\psGld,\psGld}$ are shown in \Fig{fig:feyn1}.\footnote{The one-loop corrections to the scalar potential will move the minimum from its tree-level position, generating tadpole diagrams that might contribute to $\Omega$. However, if we neglect gauge interactions, there is no $\psGld\psGld \omega^0$ coupling and tadpoles do not contribute to $\Omega$.}

Because the theory has an underlying SUSY, UV divergences cancel in the sum over the states running in the loop, so at one loop the bare quantity $\Omega_{\psGld,\psGld}$ is finite and $\widehat{\Omega}_{\psGld,\psGld}\equiv\Omega_{\psGld,\psGld}$.  The self-energy correction is
\begin{equation}
\label{eq:selfenergycorrection}
\widehat{\Omega}_{\psGld,\psGld}(p^2)=\frac{\lambda^2}{16\pi^2}\left(\sum_{k,\ell}(\mathcal{Z}^{k\ell})^2m_lB(p^2;m_\ell^2,\mu_k^2)
+2\sum_{r,s}\mathcal{Y}^{rs}\mathcal{J}^{rs}m_sB(p^2;m_s^2,\mu_r^2)\right),
\end{equation}
where $m_\ell$, $m_s$, $\mu_k$, and $\mu_r$ are the neutralino, chargino, neutral scalar, and charged scalar masses respectively, and $p$ is the external momentum.  The finite part of the (one loop) Passarino-Veltman function is
\begin{equation}
B(p^2;x,y)=-\int_0^1dt\log\left[\frac{tx+(1-t)y-t(1-t)p^2}{Q^2}\right],
\end{equation}
where the renormalization group scale $Q^2$ cancels in \Eq{eq:selfenergycorrection}.  Strictly speaking, $\widehat{\Omega}_{\psGld,\psGld}(p^2)$ should be evaluated at $\sqrt{p^2}=m_\psGld$ when used in \Eq{eq:mass2}, but since the self-energy is already $\order(\lambda^2)$, we can safely evaluate it at the tree-level mass $\sqrt{p^2}=m_\psGld^{\rm tree}\approx 0$.

\section{Dominant Annihilation Channel}
\label{app:annihilation}

\begin{figure}
\centering
\includegraphics[scale=0.7]{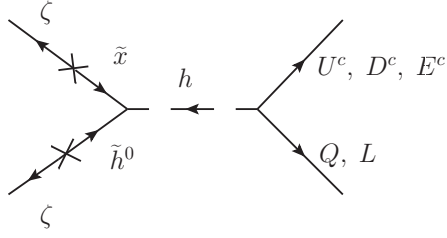}
\caption{The pseudo-goldstino can annihilate through its higgsino component to SM quarks and leptons.  However, this cross section is very small, and the pseudo-goldstino is a cosmologically dangerous hot relic.}
\label{fig:annhiliation_diagrams}
\end{figure} 

In this appendix, we confirm the statement in \Sec{sec:Cosmo} that the pseudo-goldstino is a hot relic.  At temperatures above the Higgs mass, the pseudo-goldstino has unsuppressed interactions with Higgs bosons and higgsinos.  Therefore, the pseudo-goldstino achieves thermal equilibrium with the SSM for high enough reheat temperature.  However, the interaction cross section drops rapidly for temperatures below the Higgs mass, and the freezeout temperature of the pseudo-goldstino is roughly the same as the freezeout temperature of the higgsino.

To see this, note that at temperatures below the Higgs mass, the dominant coupling of the pseudo-goldstino to light SM fields is Higgs exchange, shown in \Fig{fig:annhiliation_diagrams}.  We can estimate this cross-section as 
\be
\sigma v &\approx& \sum_f \Theta^2_{\widetilde{h}^0,\psGld} \lambda^2 y_f^2 \frac{(m_{\psGld} v)^2}{m_{h^0}^4}\left(1-\frac{4 m^2_f}{s}\right)^{3/2} \\ 
&\approx& (10^{-14}~\text{pb}) \, v^2 \parfrac{\lambda}{1.0}^2 \parfrac{y_f}{10^{-3}}^2 \left(\frac{\Theta_{\widetilde{h}^0,\psGld}}{10^{-3}}\right)^2  \left(\frac{120 \; \gev}{m_{h^0}}\right)^4 \left(\frac{m_{\psGld} }{100 \; \mev}\right)^2, \nonumber
\ee
where $y_f$ ($m_f$) is the Yukawa coupling (mass) of the relevant fermion, $m_{h^0}$ is the physical Higgs mass, $v$ is the relative velocity, and $s$ is the squared center-of-mass energy.  We have ignored the phase space suppression in the last estimate, and have used typical masses and mixing angles from \Figs{fig:PGmass}{fig:PGmixing}.  

Comparing the scattering and Hubble rates at $T=m_{\psGld}$, we have
\be
\left.\frac{\Gamma}{H} \right|_{T=m_{\psGld}} \simeq \frac{ m^3_{\psGld}\sigma v}{g_* (m^2_{\psGld}/M_{pl})} \approx  10^{-7} \parfrac{g_*}{50}\parfrac{m_{\psGld}}{100 \; \mev} \parfrac{\sigma v}{10^{-14}~{\rm pb}},
\ee
where $g_*$ is the number of degrees of freedom in equilibrium at this temperature. The scattering rate is much smaller than the Hubble rate, implying that the pseudo-goldstino freezes out while relativistic.

\section{Neutralino Decay Widths}
\label{app:Neutralinos}

\begin{figure}
\begin{center}
\subfloat[]{\includegraphics[scale=0.7]{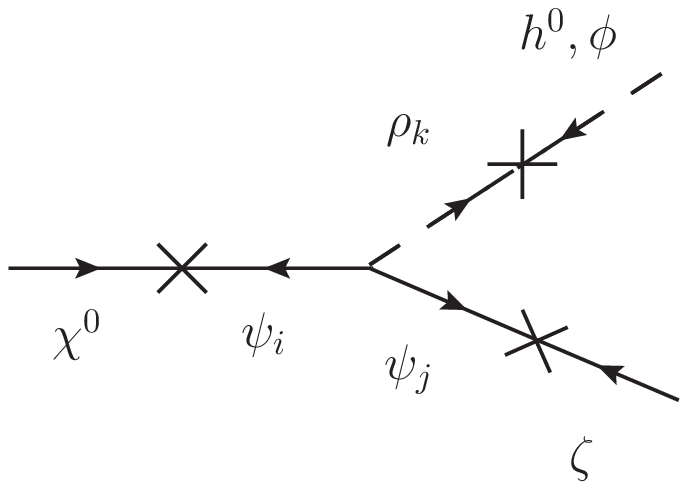}} $\qquad$
\subfloat[]{\includegraphics[scale=0.7]{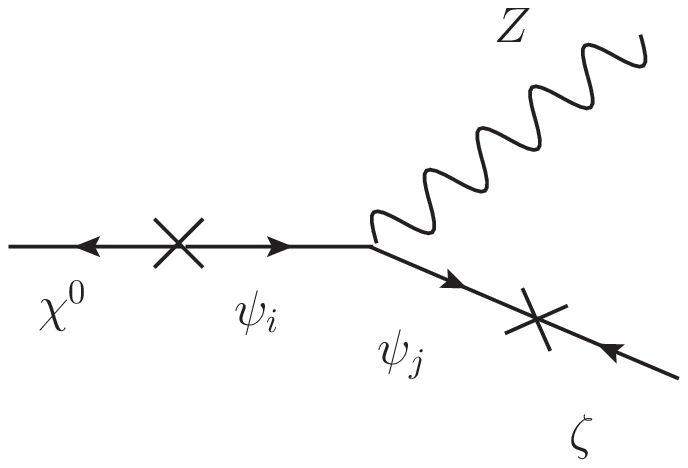}}
\end{center}
\caption{Illustrations of the general structure contributing to the neutralino LOSP decay to a pseudo-goldstino. The left figure shows the decays to scalars $h^0$ and $\phi$, while the right figure shows the decay to the $Z$.  The fermion interaction eigenstates contributing to the decay are ${\bs \psi} = \{\widetilde{x},\widetilde{h}^0_u,\widetilde{h}^0_d,\widetilde{r}^0_u,\widetilde{r}^0_d,\widetilde{B},\widetilde{W}_3\}$, and the relevant scalar interaction  eigenstates are ${\bs \rho} = \{x,h^0_u,h^0_d,r^0_u,r^0_d\}$.}
\label{fig:FullNdecays}
\end{figure}

In order to calculate the neutralino decay rates for \Sec{sec:NeutralinoDecays}, we have to account for the fact that a neutralino LOSP is in general an admixture of the higginos, gauginos, $r$-inos, and $x$-ino.  The values of the soft parameters determine the relative fractions of these components, which is especially important when the scale of $R$-violation in the visible sector is small.

The generalization of \Fig{fig:neutralinoDecays} is shown in \Fig{fig:FullNdecays}. Taking into account all of the mixings at tree-level and following the treatment in \Ref{Dreiner:2008tw}, we obtain the partial widths
\be
\Gamma(\chi^0\rightarrow Z+\psGld)&=&\frac{g^2m_{\chi^0}}{64\pi\cos^2\theta_w}|\alpha_1|^2\left(1-\frac{m_{Z}^2}{m_{\chi^0}^2}\right)\left(1-2\frac{m_{Z}^2}{m_{\chi^0}^2}\right), \\
\Gamma(\chi^0\rightarrow h^0+\psGld)&=&\frac{m_{\chi^0}}{64\pi}|\alpha_2|^2\left(1-\frac{m_{h^0}^2}{m_{\chi^0}^2}\right)^2, \\
\Gamma(\chi^0\rightarrow\phi+\psGld)&=&\frac{m_{\chi^0}}{64\pi}|\alpha_3|^2\left(1-\frac{m_\phi^2}{m_{\chi^0}^2}\right)^2,
\ee
where the relevant combinations of the mixing angles are
\be
\alpha_1&=&\Theta^{*}_{\widetilde{r}_d,\psGld}\Theta_{\widetilde{r}_d,\chi^0}-\Theta^{*}_{\widetilde{r}_u,\psGld}\Theta_{\widetilde{r}_u,\chi^0},  \\
\alpha_2&=&(\Theta_{r_u,h^0}^*\Theta_{\widetilde{r}_u,\psGld}-\Theta_{r_d,h^0}^*\Theta_{\widetilde{r}_d,\psGld}) (g'\Theta_{\widetilde{B},\chi^0}-g\Theta_{\widetilde{W},\chi^0})  \nonumber \\
 & &~ +\sqrt{2}\lambda\Theta_{\widetilde{x},\psGld}(\Theta_{\widetilde{h}_u,\chi^0}\Theta_{\widetilde{h}_d,h^0}-\Theta_{\widetilde{h}_d,\chi^0}\Theta_{\widetilde{h}_u,h^0}), \\
\alpha_3&=&(\Theta_{r_u,\phi}^*\Theta_{\widetilde{r}_u,\psGld}-\Theta_{r_d,\phi}^*\Theta_{\widetilde{r}_d,\psGld}) (g'\Theta_{\widetilde{B},\chi^0}-g\Theta_{\widetilde{W},\chi^0}) \nonumber,
\ee
and we have neglected terms that depend on the mixing of the higgsinos and gauginos with the pseudo-goldstino.  In all these expressions, the pseudo-goldstino is approximated as massless.

\bibliography{GoldstiniSSM}
\bibliographystyle{JHEP}

\end{document}